\documentclass[aps,prc,preprint,superscriptaddress,showpacs]{revtex4-1}

\usepackage{graphicx}
\usepackage{amsmath,amssymb,times}
\usepackage{ulem}

\def\fun#1#2{\lower3.6pt\vbox{\baselineskip0pt\lineskip.9pt
\ialign{$\mathsurround=0pt#1\hfil##\hfil$\crcr#2\crcr\sim\crcr}}}

\newcommand{\beq}{\begin{equation}}
\newcommand{\eeq}{\end{equation}}
\newcommand{\bea}{\begin{eqnarray}}
		  \newcommand{\eea}{\end{eqnarray}}

\DeclareSymbolFont{boldletters}{OML}{cmm} {b}{it}
\DeclareSymbolFontAlphabet{\mathbit}{boldletters}
\DeclareMathSymbol{\alpha}{\mathalpha}{letters}{"0B}
\DeclareMathSymbol{\beta}{\mathalpha}{letters}{"0C}
\DeclareMathSymbol{\gamma}{\mathalpha}{letters}{"0D}
\DeclareMathSymbol{\delta}{\mathalpha}{letters}{"0E}
\DeclareMathSymbol{\epsilon}{\mathalpha}{letters}{"0F}
\DeclareMathSymbol{\zeta}{\mathalpha}{letters}{"10}
\DeclareMathSymbol{\eta}{\mathalpha}{letters}{"11}
\DeclareMathSymbol{\theta}{\mathalpha}{letters}{"12}
\DeclareMathSymbol{\iota}{\mathalpha}{letters}{"13}
\DeclareMathSymbol{\kappa}{\mathalpha}{letters}{"14}
\DeclareMathSymbol{\lambda}{\mathalpha}{letters}{"15}
\DeclareMathSymbol{\mu}{\mathalpha}{letters}{"16}
\DeclareMathSymbol{\nu}{\mathalpha}{letters}{"17}
\DeclareMathSymbol{\xi}{\mathalpha}{letters}{"18}
\DeclareMathSymbol{\pi}{\mathalpha}{letters}{"19}
\DeclareMathSymbol{\rho}{\mathalpha}{letters}{"1A}
\DeclareMathSymbol{\sigma}{\mathalpha}{letters}{"1B}
\DeclareMathSymbol{\tau}{\mathalpha}{letters}{"1C}
\DeclareMathSymbol{\upsilon}{\mathalpha}{letters}{"1D}
\DeclareMathSymbol{\phi}{\mathalpha}{letters}{"1E}
\DeclareMathSymbol{\chi}{\mathalpha}{letters}{"1F}
\DeclareMathSymbol{\psi}{\mathalpha}{letters}{"20}
\DeclareMathSymbol{\omega}{\mathalpha}{letters}{"21}
\DeclareMathSymbol{\varepsilon}{\mathalpha}{letters}{"22}
\DeclareMathSymbol{\vartheta}{\mathalpha}{letters}{"23}
\DeclareMathSymbol{\varpi}{\mathalpha}{letters}{"24}
\DeclareMathSymbol{\varrho}{\mathalpha}{letters}{"25}
\DeclareMathSymbol{\varsigma}{\mathalpha}{letters}{"26}
\DeclareMathSymbol{\varphi}{\mathalpha}{letters}{"27}
\DeclareMathSymbol{\Gamma}{\mathalpha}{letters}{"00}
\DeclareMathSymbol{\Delta}{\mathalpha}{letters}{"01}
\DeclareMathSymbol{\Theta}{\mathalpha}{letters}{"02}
\DeclareMathSymbol{\Lambda}{\mathalpha}{letters}{"03}
\DeclareMathSymbol{\Xi}{\mathalpha}{letters}{"04}
\DeclareMathSymbol{\Pi}{\mathalpha}{letters}{"05}
\DeclareMathSymbol{\Sigma}{\mathalpha}{letters}{"06}
\DeclareMathSymbol{\Upsilon}{\mathalpha}{letters}{"07}
\DeclareMathSymbol{\Phi}{\mathalpha}{letters}{"08}
\DeclareMathSymbol{\Psi}{\mathalpha}{letters}{"09}
\DeclareMathSymbol{\Omega}{\mathalpha}{letters}{"0A}

\numberwithin{equation}{section}

\def\delsla{\!\not\!\partial}

\begin{document}
\title{Nonlocal Polyakov-Nambu--Jona-Lasinio model and imaginary chemical potential}

\author{Kouji Kashiwa}
\email[]{kashiwa@ribf.riken.jp}
\affiliation{ RIKEN BNL Research Center, Brookhaven National Laboratory, Upton,
New York 11973, USA}
\affiliation{Physik-Department, Technische Universit{\"a}t M{\"u}nchen, D-85747 Garching, Germany}
\affiliation{
Department of Physics, Graduate School of Sciences, Kyushu
University, Fukuoka 812-8581, Japan}

\author{Thomas Hell}
\email[]{thell@ph.tum.de}
\affiliation{Physik-Department, Technische Universit{\"a}t M{\"u}nchen, D-85747 Garching, Germany}

\author{Wolfram Weise}
\email[]{weise@ph.tum.de}
\affiliation{Physik-Department, Technische Universit{\"a}t M{\"u}nchen, D-85747 Garching, Germany}

\begin{abstract}
With the aim of setting constraints for the modeling of the QCD phase diagram, the phase structure of the two-flavor Polyakov-loop extended Nambu and Jona-Lasinio (PNJL) model is investigated in the range of imaginary chemical potentials ($\mu_\mathrm{I}$) and compared with available $N_f=2$ lattice QCD results.
The calculations are performed using the advanced nonlocal version of
 the PNJL model with the inclusion of vector-type quasiparticle interactions between quarks, and with wave-function-renormalization corrections.
It is demonstrated that the nonlocal PNJL model reproduces important features of QCD at finite $\mu_\mathrm{I}$, such as the Roberge-Weiss (RW) periodicity and the RW transition.
Chiral and deconfinement transition temperatures for $N_f=2$ turn out to coincide both at zero chemical potential and at finite $\mu_\mathrm{I}$.
Detailed studies are performed concerning the RW endpoint and its neighborhood where a first-order transition occurs. 
\end{abstract}

\maketitle

\section{Introduction}

One of the fundamental topics in the physics of the strong interaction is the investigation of the QCD phase diagram.
Lattice gauge theory provides a powerful tool for dealing with the thermodynamics of QCD.
However, lattice QCD (LQCD) simulations are not feasible at real values ($\mu_\mathrm{R}$) of the quark chemical potential(s), at least not for $\mu_\mathrm{R}/T>1$ at given temperature $T$, because of the well-known sign problem.
Explorations into the domain of large $\mu_\mathrm{R}$, or high baryon densities, can so far only be pursued in terms of models,  a frequently used one being the Polyakov-loop-extended Nambu and Jona-Lasinio (PNJL) model ~\cite{Meisinger,Fukushima1,Ratti,Ciminale,Sakai1,Hell1,Hell2,Kashiwa1,Abuki}.
Such approaches have obvious intrinsic limitations. (For example, important constraints well known from nuclear physics around and above normal nuclear matter densities are not (yet) realistically incorporated in such models).
Their predictive power concerning possible phase transitions at high baryon densities 
is quite limited and involves substantial ambiguities at finite $\mu_\mathrm{R}$ \cite{Kashiwa1,Abuki}. 
Setting additional constraints for the modeling of QCD phases at nonzero chemical potentials is therefore an important issue.

Under such circumstances, the continuation from real to imaginary chemical potential turns out to be a useful conceptual strategy. In this case LQCD simulations do not face the sign problem; there is no principal restriction in performing LQCD computations in the region of imaginary chemical potentials. 
Models formally designed for general complex chemical potentials, 
\begin{align}
\mu = \mu_\mathrm{R} +
 i\mu_\mathrm{I}~,\nonumber
\end{align}
can then interpolate between positive and negative $\mu^2$ and incorporate LQCD constraints set at imaginary $\mu$ into extrapolations to the physical region of real $\mu$.

The relationship between $\mu_\mathrm{R}$ and $\mu_\mathrm{I}$ is apparent in the connection between the grand canonical partition function, $Z_\mathrm{GC}(T,\theta)$, written for imaginary chemical potential in terms of the dimensionless ratio $\theta=\mu_\mathrm{I}/T$, and the canonical partition function $Z_\mathrm{C} (T,N_q)$ expressed as a function of quark number $N_q$\,:
\begin{align}
Z_\mathrm{C} (T,N_q) 
&= \int^\pi_{-\pi} \frac{ \mathrm{d} \theta }{ 2 \pi } 
   e^{-iN_q\theta} Z_\mathrm{GC}(T,\theta)~.
\end{align}
Inverting this Fourier transform gives the grand partition function at finite $\mu_\mathrm{R}$ through the fugacity expansion as
\begin{align}
Z_\mathrm{GC} (T,\mu_\mathrm{R})
&= \sum_{N=-\infty}^{\infty} e^{N \mu_\mathrm{R}/T} Z_\mathrm{C} (T,N)~,
\end{align}
where the thermodynamical limit is considered, with the three-dimensional volume set to $V \to \infty$.

Several important features are noteworthy in the $\mu_\mathrm{I}$ region (see Refs.~\cite{Sakai1,Sakai2,Kouno,Kashiwa2,Kashiwa3,RW} for details). 
First, the QCD partition function has a $2\pi/3$ periodicity along the $\theta$ axis, the Roberge-Weiss (RW) periodicity~\cite{RW}.
Another characteristic property is the RW transition: thermodynamical quantities and order parameters have nonanalytic behavior at $\theta = \pi k/3$ (with $k$ an odd integer) once the temperature exceeds a specific value, $T\geq T_\mathrm{E}$. 
This point, $(T,\theta)=(T_\mathrm{E},\pi k/3)$, is the so-called RW endpoint.
These nonanalyticities are induced by the transition between different minima of the thermodynamical potential (see Fig.~2 in Ref.~\cite{RW}). 
Moreover, the imaginary chemical potential is gauged into the temporal boundary condition of quarks; the dual quark condensate~\cite{Bilgici,Kashiwa4,Fischer,Braun,Mukherjee,Gatto} is then proposed as a measure for exploring the possible correlation between the chiral and the deconfinement transitions.
This correlation was also studied previously in terms of a Ginzburg-Landau analysis~\cite{Mocsy}.

In recent years several LQCD investigations at imaginary chemical potentials have been performed in order to draw conclusions about the $\mu_\mathrm{R}$ region, either through analytical continuation~\cite{Forcrand1,D'Elia,Chen,Wu,Forcrand2} or by the canonical approach~\cite{Forcrand3,Li}.
These studies were so far restricted to a small range of $\mu_\mathrm{R}$.
To overcome such limitations, the imaginary chemical potential matching approach has been proposed in Ref.~\cite{Kashiwa3}.
The basic aim of this approach is the modeling of the QCD phase diagram starting from a suitable effective Lagrangian constrained by direct comparison with LQCD results in the region of imaginary chemical potentials $\mu_\mathrm{I}$.
The implementation of the correct QCD behavior in the $\mu_\mathrm{I}$ region is then a necessary (though not sufficient) requirement for proceeding to physical (real) chemical potentials.

The analytic continuation of LQCD data from $\mu_\mathrm{I}$ to $\mu_\mathrm{R}$ commonly assumes a certain form (e. g. polynomial) for the interpolating function. This implies a limiting convergence radius. (See, for example, Ref.~\cite{Cea1,Cea2} for a detailed assessment).  The strategy pursued in the present work is different. Given an effective Lagrangian starting from QCD and its symmetries, a set of possible types of effective interactions between quarks is introduced. To the extent that these interactions are not entirely determined just by the physical low-mass meson spectrum, lattice QCD results at finite $\mu_\mathrm{I}$ are useful in providing additional constraints. Once the relevant effective interactions are fixed, the continuation to finite $\mu_\mathrm{R}$ is not subject to convergence radius limitations any more.

The PNJL model is of considerable interest in this context.
In previous work~\cite{Sakai1} it was shown that the PNJL model can reproduce the RW periodicity together with other important QCD properties such as spontaneous chiral symmetry breaking.
The RW periodicity is described by the extended ${\mathbb Z}_3$ symmetry defined as
\begin{align}
\theta \to \theta + 2\pi n/3~,\qquad
\Phi \to e^{-2\pi i n/3}\Phi~,\qquad
{\bar \Phi} \to e^{2\pi i n/3}{\bar \Phi}~,
\end{align}
where $n$ is any integer; $\Phi$ and ${\bar \Phi}$ are the Polyakov loop and its conjugate, respectively.
In the PNJL model this symmetry is realized by construction.
It has in fact been shown that the {\it local} version of the PNJL model is capable of reproducing the LQCD data at finite $\mu_\mathrm{I}$, but at the expense of introducing  scalar-type eight-quark and vector-type four-quark interactions~\cite{Sakai2}, or a Polyakov-loop dependent NJL coupling strength~\cite{Sakai3}, in addition to the standard (scalar plus pseudoscalar) chiral contact interactions between quarks.

The local PNJL model still works with a schematic ad-hoc momentum-space cutoff that has no foundation in QCD.
More direct contacts with QCD have recently been established by introducing the {\it nonlocal} PNJL model~\cite{Hell1,Hell2,Contrera,Hell3} and by pointing out its formal derivation from QCD~\cite{Kondo}.
 Nonlocal interactions were also investigated, for example, within the framework of NJL ~\cite{Langfeld} and  instanton models~\cite{Nam}.
It is instructive to extend the nonlocal PNJL approach to imaginary chemical potentials in order to examine its properties in direct comparison with LQCD results. Very recently, independent related calculations have been performed \cite{Pagura} in parallel to the present investigations, leading to qualitatively similar conclusions. The additional new element in the present work is the detailed study of the role of (nonderivative) vector interactions between quarks as they emerge from their basic color-current couplings.

In the present work we restrict ourselves to the two-flavor case.
Section~II briefly summarizes the nonlocal $N_f=2$ PNJL model in its
latest version~\cite{Hell3} with the inclusion of quark wave-function-renormalization effects.
Numerical results are presented and discussed in Sec.~III. The paper closes with a summary and conclusions in Sec.~IV.


\section{Framework and formalism}

This section proceeds in several steps, starting with a brief exhibition of the nonlocal PNJL model and its thermodynamics, followed by some symmetry considerations. Finally, a possible additional vector-current interaction between quarks will be incorporated.
Such a Lorentz-vector interaction is usually not part of the standard PNJL model.
It is expected to play a pronounced role, however, in studies extended to imaginary chemical potentials.   

\subsection{Nonlocal PNJL model}
\label{PNJL-q}

The generic Euclidean action of the two-flavor PNJL model is
\begin{align}
{\cal S} = \int_0^\beta\mathrm{d}\tau\int \mathrm{d}^3 x \left[{\bar q}(x) (i \!\not\!\!D - m_0 ) q(x) + {\cal L}_\mathrm{int}\right]
             - \beta V\,{\cal U}(\Phi[A],{\bar \Phi}[A];T)~,
             \label{Lagrangian}
\end{align}
where $q(x) = (u(x), d(x))^T$ is the two-flavor quark field, $m_0$ denotes the current quark mass taken in the isospin limit $(m_0 \equiv m_u = m_d)$, and $D^\nu=\partial^\nu + i A^\nu=\partial^\nu
+i\delta^{\nu}_{0}\,A^{0,a}{\lambda_a / 2}$ is the color gauge-covariant derivative, with $\mathrm{SU}(3)_c$ Gell-Mann matrices $\lambda_a$. The gauge coupling $g$ is understood to be absorbed in the definition of $A^{0,a}$.
The last term in Eq.~(\ref{Lagrangian}) is the Polyakov-loop-effective potential ${\cal U}$, multiplied by volume $V$ and inverse temperature $\beta = T^{-1}$, and to be specified later.
As in previous studies we treat the temporal gauge field $A^0$ as a constant Euclidean background field in the form $A_4 = i A^0 = A_4^3\,\lambda_3/2 + A_4^8\,\lambda_8/2 $. 
Further details are given in Sec.~\ref{PLP}.
For extensions beyond the mean-field treatment of $A_4$, see Ref.~\cite{Roessner,Cristoforetti}.
The nonlocal generalization of the PNJL model is characterized by an interaction Lagrangian featuring nonlocal quark currents and densities, as follows \cite{Hell1,Hell2,Contrera,Hell3,Pagura}: 
\begin{align}
{\cal L}_\mathrm{int} (x) 
&= G\,\Bigl[
j_\mathrm{a} (x) j_\mathrm{a} (x) 
+ J (x) J(x) \Bigr]~, \\
j_\mathrm{a} (x) 
&= \int \mathrm{d}^4z~ \tilde{\cal C}(z)~{\bar q}( x + z/2) ~\Gamma_\mathrm{a} ~
                  q( x-z/2)~,
 \label{eq6}\\
J (x)
&= \int \mathrm{d}^4z~ \tilde{\cal F}(z)~ {\bar q}(x + z/2)~
   \frac{i\delsla^{^{^{\!\!\!\!\!\leftrightarrow}}}}{2 \kappa}
                   ~q(x - z/2)~.
                      \label{eq7}
\end{align}
The chiral (scalar and pseudoscalar) densities $j_\mathrm{a}(x)$ with $\mathrm{a} = 0,1,2,3$ involve the operators $\Gamma_\mathrm{a}=(1,i\gamma_5 {\vec \tau})$. The overall coupling strength $G$ of dimension $[\text{length}]^2$ is chosen sufficiently large so that spontaneous chiral symmetry breaking and pions as Goldstone bosons emerge properly. The $J(x)$ introduces additional vector-type derivative couplings with
$~{\bar q}(x')\,\partial^{^{^{\!\!\!\!\!\leftrightarrow}}}_\mu\,q(x)
:= {\bar q}(x') (\partial_\mu q) (x) - (\partial_\mu {\bar q})(x')\,
q(x)$ together with a scale $\kappa$ so that the effective strength of
this term in ${\cal L}_\mathrm{int}$ is $G/\kappa^2$. In the following
we refer to the interaction induced by $J(x)$ simply as a {\it derivative coupling} in order to avoid confusion with the {\it nonderivative vector interaction} that will be introduced in Sec.~\ref{Sec:Vector}.

Associated with the currents or densities (\ref{eq6}) and (\ref{eq7}) are nonlocality distributions $\tilde{\cal C}(z)$ and $\tilde{\cal F}(z)$. As described in detail in Refs.~\cite{Hell1,Hell2,Contrera,Hell3}, these distributions govern the momentum dependences of the mass function and of the renormalization factor that  appear in the quark quasiparticle propagator, $Z(p^2)(\gamma\cdot p - M(p^2))^{-1}$.  The Fourier transform ${\cal C}(p^2)$ of $\tilde{\cal C}(z)$ is related to the quasiparticle mass function $M(p^2)$ determined by the self-consistent gap equation,
\begin{align}
M(p^2) &= Z(p^2) 
\Bigl[ m_0 + \sigma\,{\cal C}(p^2) \Bigr]~~,
\end{align}
where $\sigma$ is the scalar mean field basically representing the chiral (quark) condensate 
$\langle\bar{q}q\rangle$.  The Fourier transform ${\cal F}(p^2)$ of $\tilde{\cal F}(z)$ is in turn related to 
the $Z$ factor representing quark wave-function renormalization,
\begin{align}
Z(p^2) &= \Bigl[ 1 - \frac{v}{\kappa}\, {\cal F}(p^2) \Bigr]^{-1}~~,
\end{align} 
where $v$ is the mean field induced by $J(x)$~\cite{Noguera}.

\subsection{Nonlocality distributions, quark mass function, and quasiparticle renormalization factor} 

The following momentum-space forms of the distribution functions appearing in Eqs.\,(\ref{eq6}) and (\ref{eq7}) are used in the present work:
\begin{align}
{\cal C} (p^2) &= \int \mathrm{d}^4z \,\exp(-i p\cdot z)\, \tilde{\cal C}(z)
= \begin{cases}
e^{-p^2 d_C^2/2} & (p^2< \lambda^2)\\
{\cal N}\,\frac{\alpha_s(p^2)}{p^2}&(p^2 \geq \lambda^2)~,\end{cases}
\label{eq10}
\\
{\cal F}(p^2) &= \int \mathrm{d}^4z \,\exp(-i p\cdot z)\, \tilde{\cal F}(z) =  \exp \Bigl( -p^2 d_F^2/2 \Bigr)~.
\label{eq:form_factor}
\end{align}
The running QCD coupling $\alpha_s(p^2)$ determines the asymptotic form of ${\cal C}(p^2)$ while its infrared behavior is given a Gaussian parametrization with a characteristic length scale $d_C$.
The matching of these high- and low-momentum representations at an intermediate scale $\lambda$ determines the constant ${\cal N}$.
The distribution ${\cal F}(p^2)$ has its own length scale $d_F$ over which the nonlocality unfolds. As in Ref.~\cite{Hell3} we use $d_C \approx 0.4$ fm and $d_F \approx 0.3$ fm. 

\begin{figure}[htbp]
\begin{center}
 \includegraphics[width=0.45\textwidth]{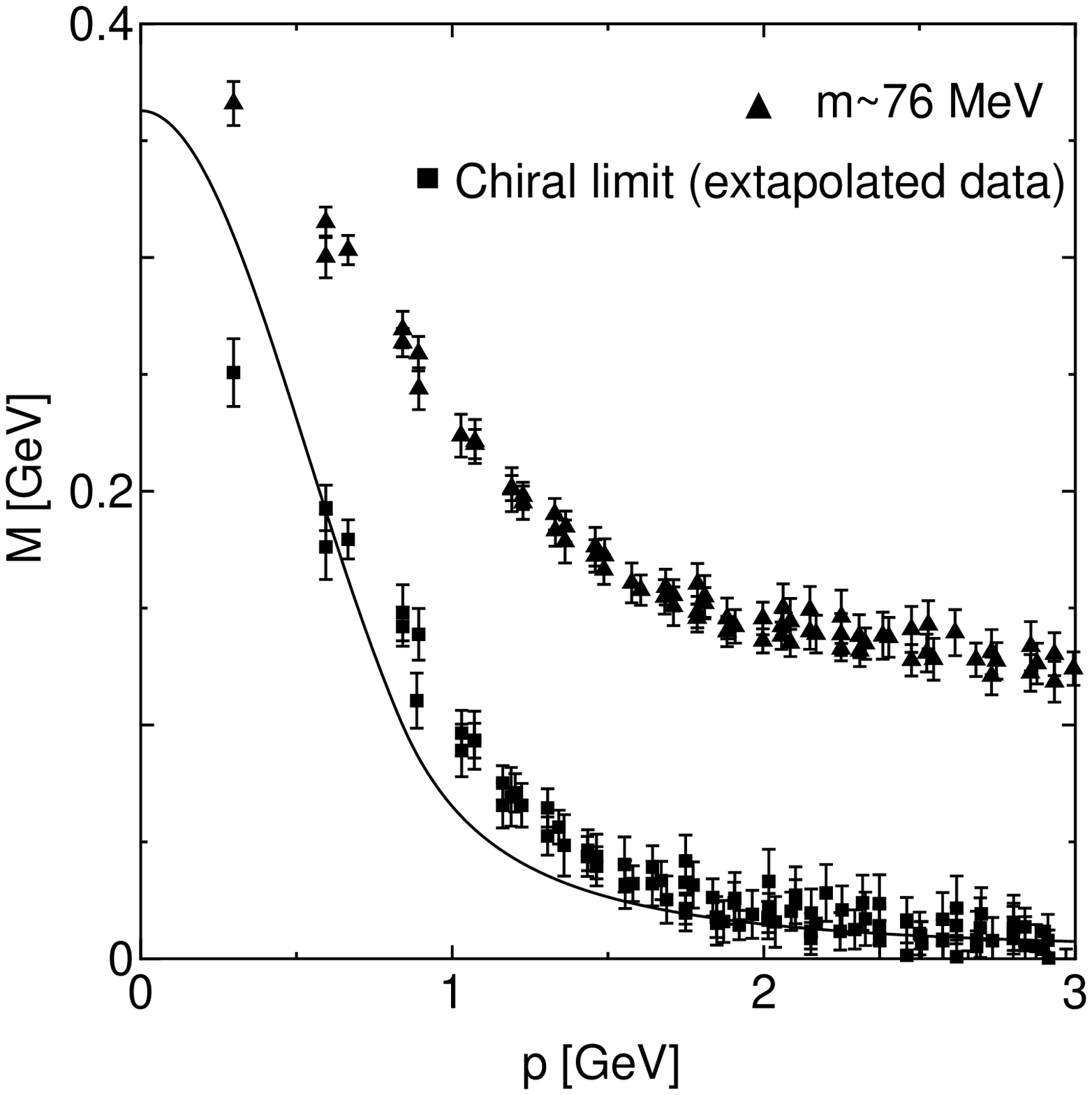}\hfill
 \includegraphics[width=0.45\textwidth]{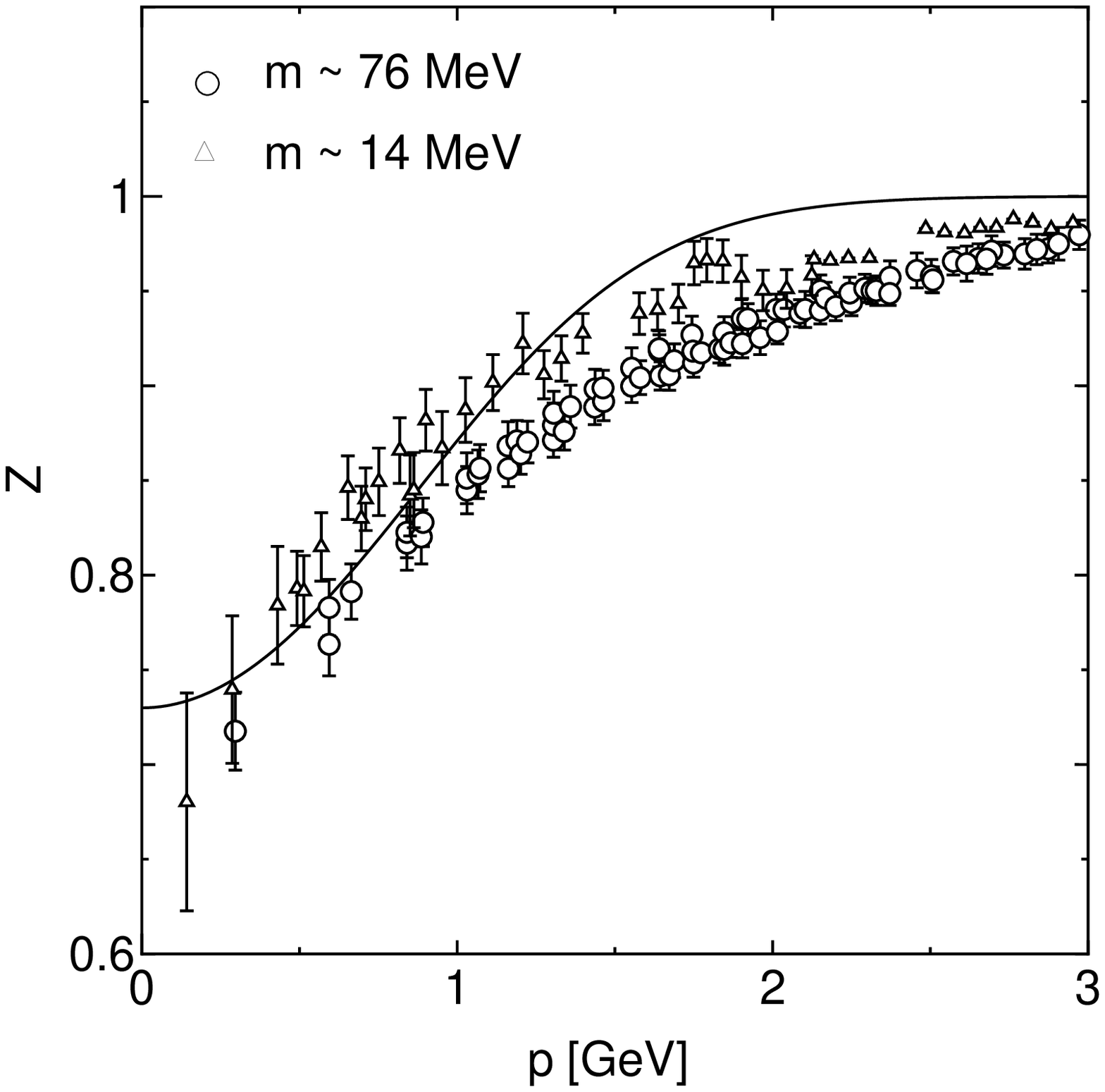}
\end{center}
\vspace{-.5cm}
\caption{The $p$ dependence of the quark mass function $M(p^2)$ and of the quasiparticle renormalization factor $Z(p^2)$ resulting from the distributions (\ref{eq10}) and (\ref{eq:form_factor}) (solid lines). Solid triangles and open circles are LQCD data generated with a large quark mass, while solid squares show the extrapolation of the mass-function data to the chiral limit (from Ref.~\cite{Bowman}). Open triangles are LQCD results for $Z(p^2)$ taken from Ref.~\cite{Parappilly}.
}
\label{Fig:Mass_Z-SC}
\end{figure}

With this input, the resulting mass function $M(p^2)$ and the renormalization factor $Z(p^2)$ closely resemble LQCD results, as shown in Fig.~\ref{Fig:Mass_Z-SC}. For the mass function we are guided by the lattice data of Ref.~\cite{Bowman} extrapolated to the chiral limit $(m_0 \rightarrow 0)$. Both the mass function and the $Z$ factor are gauge dependent.  Using Landau and Laplacian gauges in comparison, it is demonstrated in 
\cite{Bowman} that the mass function shows very little variation between these two gauge fixings, whereas the gauge dependence of the $Z$ factor is about 20 \% in the infrared region. This gauge dependence of $Z$ is not crucial here since the wave-function-renormalization effects will turn out to be quite small in the present context. In practice we use the Landau-gauge results for orientation. 

At finite temperature $T$ the squared (Euclidean) four-momentum in ${\cal C}(p^2)$ and ${\cal F}(p^2)$ becomes $p^2 = \omega_n^2 + \vec{p}\,^2$  with fermionic Matsubara frequencies $\omega_n = (2n + 1)\pi T$. When Polyakov-loop fields and nonzero chemical potentials are included,  $\omega_n$ is shifted to $\omega_n -i\mu$ plus linear combinations of $A_4^{3,8}$.

In the PNJL model the pion mass and its decay constant are used to fit parameters in the NJL sector of the Lagrangian. 
These parameters are taken from Refs.~\cite{Hell1,Hell3} for the case
studies without and with the inclusion of $Z$-factor effects, respectively. 

\subsection{Polyakov-loop-effective potential} 
\label{PLP}

The Polyakov-loop-effective potential ${\cal U}$ is used in the form given in Ref.~\cite{Hell1}: 
\begin{align}
{{\cal U} ({\bar \Phi},\Phi;T)\over T^4}
&= -\frac{1}{2}\, b_2(T)\, {\bar \Phi}\, \Phi
              + b_4(T)\, \ln[ 1 - 6\, {\bar \Phi} \,\Phi 
                            +4 ({\bar \Phi}^3 + \Phi^3) 
                             -3 ({\bar \Phi}\,\Phi)^2]~, 
\end{align}
where $\Phi$ and ${\bar \Phi}$ are represented as
\begin{align}
\Phi 
&= \frac{1}{3} \left[\exp\left(i {A_4^3 + A_4^8\over 2T}\right) + \exp\left(-i {A_4^3 - A_4^8\over 2T}\right)  + \exp\left(i {A_4^8\over \sqrt{3}T}\right) \right]~,
\qquad
{\bar \Phi} 
 = \Phi^*~.
\label{eq:polyakovloop}
\end{align}
The coefficient functions $b_2(T)$ and $b_4(T)$ are parametrized to reproduce pure-gauge LQCD results as described in Refs.~\cite{Hell1,Hell3}. 
The temperature scale $T_0$ appearing in $b_2(T)$ and $b_4(T)$ is set to $270\,\text{MeV}$, the critical temperature for the first-order confinement-deconfinement transition from LQCD in the pure-gauge limit.
Variations of this scale in the presence of dynamical quark flavors \cite{Schaefer} are of potential importance but will not be considered here. 

\subsection{Thermodynamics}

The mean-field thermodynamical potential $\Omega$ of the nonlocal PNJL model, including quark wave-function-renormalization corrections, is constructed using the Nambu-Gor'kov formalism (see Ref.~\cite{Meif} and references therein). The final form is
\begin{align}
\Omega &=
\Omega_1 + \Omega_{\rm free} + {\cal U}(\Phi,{\bar \Phi};T)~,
\label{TP-PNJL}
\end{align}
where
\begin{align}
&\Omega_1 
= - 4T \sum_{i=\pm,0} \sum_{n=-\infty}^{\infty}
           \int \frac{\mathrm{d}^3 p}{(2\pi)^3}
           \ln \left[ \frac{ \omega_{n,i}^2 + E_i^2(p^2) }
                    { (\omega_{n,i}^2 + \vec{p}\,^2 + m_0^2 )~Z_i^2 (p^2) } 
     \right]
+  \frac{ \sigma^2 + v^2}{4G}~.
\label{Eq:OT}
\end{align}
Here $\sigma$ and $v$ are the mean fields associated with the scalar density $j_0$ of Eq.~(\ref{eq6}) and the derivative-vector current $J$ of Eq.~(\ref{eq7}), respectively, upon bosonization. The first term on the right-hand-side of Eq.~(\ref{Eq:OT}) involves the quark quasiparticle energies
\begin{align}
E_i &= \sqrt{\vec{p}\,^2+M_i^2(p^2)}
\end{align}
with dynamically generated masses, $M_i(p^2) \equiv M (p^2 = \omega_{n,i}^2 + \vec{p}\,^2)$, determined self-consistently at each shifted Matsubara frequency $\omega_{n,i}$ with $i \in \{ \pm,0\}$:
\begin{align}
\omega_{n,\pm} &= \omega_n  - i \mu \pm \frac{A_4^3}{2} -
 \frac{A_4^8}{2\sqrt{3}}~~,\nonumber \\
\omega_{n,0  } &= \omega_n - i \mu + \frac{A_4^8}{\sqrt{3}}~,
\label{eq17}
\end{align}
where $A_4^{3,8}$ are the gauge fields forming the Polyakov loop already given in Eq.~\eqref{eq:polyakovloop}.
Likewise, the $Z$ factors are understood as $Z_i(p^2)\equiv Z(p^2=\omega_{n,i}^2+\vec{p}\,^2)$.
More explicitly:
\begin{align}
M_i(p^2) &= Z_i(p^2) 
\Bigl[ m_0 + \sigma\,{\cal C}(p^2=\omega_{n,i}^2+\vec{p}\,^2) \Bigr]~,\\
Z_i(p^2) &= \Bigl[ 1 - \frac{v}{\kappa}\, {\cal F}(p^2=\omega_{n,i}^2+\vec{p}\,^2) \Bigr]^{-1}~.
\end{align}

At finite temperature $T$ the Lorentz invariance is broken by the thermal medium and the inverse quark quasiparticle propagator becomes $S^{-1}(p) = {\cal A}_0(p)\, \gamma_0 p^0 + {\cal A}(p) \gamma_i p^i + {\cal B}(p)$ with ${\cal A}_0 \neq {\cal A}$. Here we assume for simplicity that the difference between ${\cal A}_0$ and ${\cal A}$ is sufficiently  small so that it can be neglected, given that the overall influence of wave-function renormalization on thermodynamical quantities is not very significant.

The subtracted $\ln(\omega_{n,i}^2 + \vec{p}\,^2 + m_0^2)$ piece in Eq.~(\ref{Eq:OT}) improves convergence in the summation over the Matsubara frequencies. The term $\Omega_\mathrm{free}$ in Eq.~(\ref{Eq:OF}),
\begin{align}
&\Omega_\mathrm{free} 
= -4 T \int \frac{\mathrm{d}^3 p}{(2\pi)^3}  
   \Bigl[ \ln \Bigl(1+3(\Phi
                     +{\bar \Phi}e^{-\beta e_-(\vec{p}\,)})
                      e^{- \beta e_-(\vec{p}\,)} 
                     +e^{- 3\beta e_-(\vec{p}\,)} \Bigr) 
\nonumber\\
&\hspace{3.5cm}  +\ln \Bigl(1+3({\bar \Phi}
                     +\Phi e^{- \beta e_+(\vec{p}\,)})
                      e^{- \beta e_+(\vec{p}\,)} 
                     +e^{- 3\beta e_+(\vec{p}\,)} \Bigr)
\Bigr]~,
\label{Eq:OF}
\end{align}
(with $\beta = 1/T$) is then introduced for consistency. 
In these ``free'' parts the quark energies, shifted by the chemical potential $\mu$, are taken with the current quark mass $m_0$:
\begin{align}
e_{\pm}(\vec{p}\,) &= \sqrt{\vec{p}\,^2+m_0^2} \pm \mu~.
\label{eq21}
\end{align}
The mean fields $\sigma$ and $v$ are determined by the conditions
\begin{align}
\frac{\partial \Omega}{\partial \sigma}=\frac{\partial \Omega}{\partial v}=0~. 
\end{align}
Similarly, the Polyakov-loop-background gauge fields are determined through
\begin{align}
\frac{\partial \Omega}{\partial A_4^3}=\frac{\partial \Omega}{\partial A_4^8}=0~.
\end{align}

Fluctuations beyond the mean field - not considered in the present work - are of course important but do not change the basic phase transition pattern in a qualitative way. Studies of the role of selected fluctuations in the PNJL model have been performed previously \cite{Roessner,Cristoforetti}. See also 
Refs.~\cite{Herbst,Skokov,Morita} for the treatment of fluctuations in the Polyakov-loop extended quark-meson model.

\subsection{Symmetry considerations}

The QCD partition function is known to be a function of $\mu^2$ (see for examples Refs.~\cite{Forcrand1,Kratochvila}) as a consequence of time reversal or CP symmetry which implies invariance under the transformation $\mu \to -\mu$. Our nonlocal PNJL model maintains this property.
For later convenience, we introduce a modified Polyakov loop and its conjugate, $\Psi$ and ${\bar \Psi}$, as
\begin{align}
\Psi &= e^{i\theta} \Phi,~~~~{\bar \Psi} = e^{-i\theta} {\bar \Phi}~.
\label{eq:polyakov}
\end{align}
These quantities are invariant under extended ${\mathbb Z}_3$ transformations and hence they are RW periodic.
Properties with respect to charge-conjugation ($C$) symmetry are of basic interest in this context. As Roberge and Weiss~\cite{RW} have shown, the RW periodicity of the QCD partition function implies that the $C$ symmetry is also intact at $\theta=\pi/3$ modulo $2\pi/3$. The $\theta$-even thermodynamical potential transforms under $C$ as
\begin{align}
\Omega (\theta) \to \Omega (-\theta)~.
\end{align}
Therefore the $C$ symmetry holds at $\theta = (2n-1) \pi/3$ with integer $n$. On the other hand, quantities that 
are $\theta$-odd transform under $C$ as
\begin{align}
{\cal O}(\theta) \to - {\cal O}(-\theta)~.
\end{align}
Spontaneous breaking of $C$ symmetry is indicated when ${\cal O}$
has a nonzero expectation value at $\theta = (2n-1)\pi/3$, associated with ${\mathbb Z}_2$-symmetry breaking under $\mu$-reflection, $\mu \leftrightarrow -\mu$. This situation is realized if the quantity in question has a nonanalyticity along the $\theta$-axis, induced by the RW transition. Examples displaying this behavior are the quark number density $n_q$ and the imaginary part $\mathrm{Im}~\Psi$ of the modified Polyakov loop \cite{Kouno}.
The $C$ symmetry is explicitly broken whenever $\theta\neq 0$ and $\theta \neq(2n-1)\pi/3$, but close to the RW endpoint this symmetry can still be regarded as  approximate.

This discussion clarifies that the RW transition lines are first order, but it does not identify the order of the transition right at the RW endpoint: both first- and second-order transitions are possible at that point.
In our nonlocal PNJL model, the first-order transition will indeed be
shown to proceed on to the RW endpoint. In this case, the RW endpoint
can become a triple point at which three first-order-transition lines
meet. However, as $\theta$ moves away from $(2n-1)\pi/3$, the explicit
$C$-symmetry breaking takes over and the first-order phase transition
turns into a crossover.

It has been suggested that the deconfinement crossover at
$\mu=0$ can be considered as the remnant of the $C$-symmetry breaking (see Ref.~\cite{Kouno} for details).
As already mentioned, one can understand from this viewpoint that the transition behavior at finite $\mu_\mathrm{I}$ is connected to the chiral and  deconfinement transitions at finite $\mu_\mathrm{R}$.  Investigations of the $\mu_\mathrm{I}$ region have indeed attracted considerable attention lately.
Much recent progress has been achieved by analyzing the RW endpoint in LQCD simulations with two- and three-flavor quarks as reported in Ref.~\cite{D'Elia2,Bonati} and Ref.~\cite{Forcrand2}.
(We mention, in passing, that the RW endpoint is also investigated within the frame of gauge/string duality~\cite{Aarts}).

\subsection{Nonderivative vector couplings}
\label{Sec:Vector}

The standard NJL and PNJL approaches usually work with a chirally
symmetric combination of scalar and pseudoscalar quark couplings as a
minimal setup. On the other hand, it is well known that additional
vector and axial-vector couplings also emerge as parts of the chirally
symmetric effective interactions between quark quasiparticles. Here we
shall focus on the role of isoscalar-vector interactions and their
effects at imaginary chemical potentials. In the mean-field approximation this isoscalar-vector interaction is directly related to nonzero baryon density; it appears independently of the derivative coupling already mentioned [the one that generates the quark wave-function-renormalization factor $Z(p)$]. As elaborated in detail in the Appendix, such a nonderivative vector interaction arises naturally by Fierz transformation from the interaction between quark color currents that has in turn its origin directly in QCD.

The actual form of the additional nonlocal vector interaction used here is
\begin{align}
\delta{\cal L}_{\mathrm{int}} 
&= -G_\mathrm{v}\, j^\mu(x)\, j_\mu(x),\label{eq28}\\
j^\mu(x) 
&= \int \mathrm{d}^4z~ {\cal C}(z)\, {\bar q}( x + {z}/{2})\, \gamma^\mu \,
                  q( x-{z}/{2})~. 
                  \label{vector}
\end{align}
Introducing a mean field $\omega$ associated with the Euclidean time component (density) of the current (\ref{vector}) after bosonization, the primary effect of this interaction is to shift the chemical potential  in Eqs. (\ref{eq17}) and (\ref{eq21}) as
\begin{align}
\mu &\to  
\mu -  \omega~, 
\end{align}
while the thermodynamic potential receives an extra contribution
\begin{align}
{\Omega} &\to {\Omega} 
 - {\omega^2\over 4 G_\mathrm{v}}~.
\end{align}
The coupling strength $G_\mathrm{v}$ of the vector interaction is conveniently expressed in terms of the ratio
$G_\mathrm{v}/G$, relative to the scalar-pseudoscalar coupling $G$ that controls spontaneous chiral symmetry breaking. In the actual calculations this ratio will be allowed to vary within $0.25\le G_\mathrm{v}/G \lesssim 0.5$, corresponding to limiting axial $\mathrm{U}(1)_\mathrm{A}$ anomaly scenarios as discussed in detail in the Appendix. 

Note that the nonlocality distributions  ${\cal C}(p^2)$ and ${\cal F}(p^2)$ are introduced at the Lagrangian level; hence they are to be calculated with chemical potentials $\mu$ {\it not} shifted by the vector mean field $\omega$. Note, furthermore, that the thermodynamical potential must be real, so extensions to imaginary chemical potential imply imaginary $\omega$ at the same time~\cite{Sakai2}.

\section{Results}

As we proceed to the actual calculations, the following versions of the nonlocal PNJL model are used in comparison: 
\begin{itemize}
\item{Set I: without a $Z$ factor [i.e., $Z(p^2)\equiv 1$] and setting $G_\mathrm{v} = 0$.} 
\item{Set II: $Z\equiv 1$ but $G_\mathrm{v} \ne 0$.}
\item{Set III: $Z(p^2)$ included but $G_\mathrm{v}= 0$.}
\item{Set IV: $Z(p^2)$ and $G_\mathrm{v} \ne 0$ both included.}
\end{itemize}

This section first presents and discusses numerical results at imaginary chemical potential using the nonlocal PNJL model in its versions with sets I and II, i.e., setting the wave-function-renormalization factor $Z\equiv 1$ and comparing scenarios without and with vector interactions. In the second part, using sets III and IV, it will be demonstrated that additional effects induced by the $Z(p^2)$ factor are indeed individually small but can produce nonnegligible effects in the $T$-$\theta$ phase diagram when combined with vector couplings.  

\begin{figure}[htbp]
\begin{center}
 \includegraphics[width=0.5\textwidth]{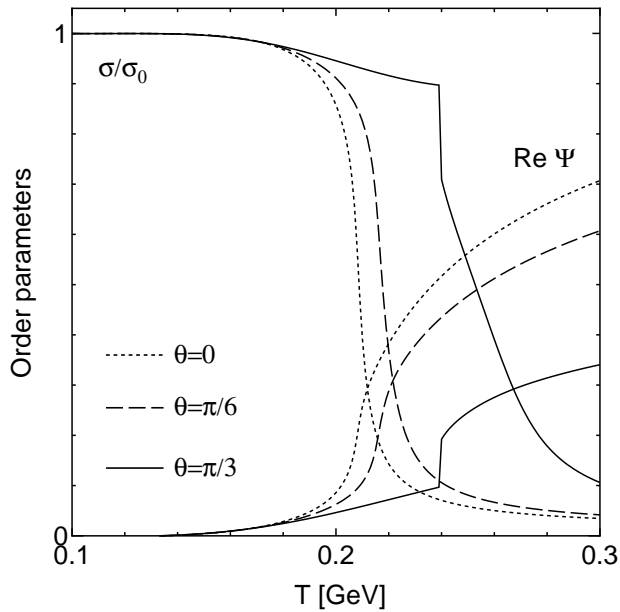}
\end{center}
\vspace{-.5cm}
\caption{ The $T$ dependence of the scalar mean field $\sigma$ and of the real part $\mathrm{Re}~\Psi$ of the modified Polyakov loop.
The dotted, dashed and solid lines denote PNJL results at $\theta = 0$, 
$\pi/6$ and $\pi/3$, respectively, using input set I.
}
\label{Fig:order_parameters_Tdep}
\end{figure}

Figure~\ref{Fig:order_parameters_Tdep} shows the temperature dependence of the scalar mean field $\sigma$, and of the real part of the modified Polyakov loop (\ref{eq:polyakov}), at different values of the 
imaginary chemical potential:  $\theta=\mu_\mathrm{I}/T = 0,~\pi/6$ and $\pi/3$. The transition temperature at $\theta=0$ is $T_c = 208\,\text{MeV}$.
At $\theta=0$ and $\pi/6$, the $\sigma$ field and $\mathrm{Re}~\Psi$ show the typical features of a crossover. 
At $\theta=\pi/3$ these quantities have already developed discontinuities, displaying gaps in both $\sigma$ and $\mathrm{Re}~\Psi$ at $T=240\,\text{MeV}$. This means that the RW endpoint becomes a first-order transition point  associated with spontaneous $C$-symmetry breaking. This feature repeats itself when $\theta$ is shifted to $\theta + 2\pi n /3$, reflecting the RW periodicity. Our results at this point are consistent with LQCD data using finite-size scaling and two-flavor staggered quarks~\cite{D'Elia2,Bonati}.

The transition temperatures for the chiral and deconfinement crossovers, $T_c^\chi$ and $T_c^{d}$, have been determined at the maximum slopes of the order parameters.  It is found that these pseudocritical temperatures coincide both at $\mu=0$ and at finite $\mu_\mathrm{I}$. This implies that two crossover lines are connected at the RW endpoint. Both lines can be considered as remnants of spontaneous $C$-symmetry breaking at $\theta=\pi/3$.
This interesting property is specific to the {\it nonlocal} PNJL
model. In local versions of this model, with an artificial cutoff in momentum space, $T_{c}^\chi$ and $T_{c}^d$ may differ. In this case the chiral crossover line does not connect with the first-order RW endpoint, and the chiral transition cannot be associated with a remnant of spontaneous $C$-symmetry breaking. Such a feature is also observed in the chiral limit at finite $\mu_\mathrm{I}$. In this case the chiral transition is second order and the transition line terminates at a point different from the RW endpoint; see for example Fig.\,1 in Ref.~\cite{Kashiwa3}.

\begin{figure}[htbp]
\begin{center}
 \includegraphics[width=0.5\textwidth]{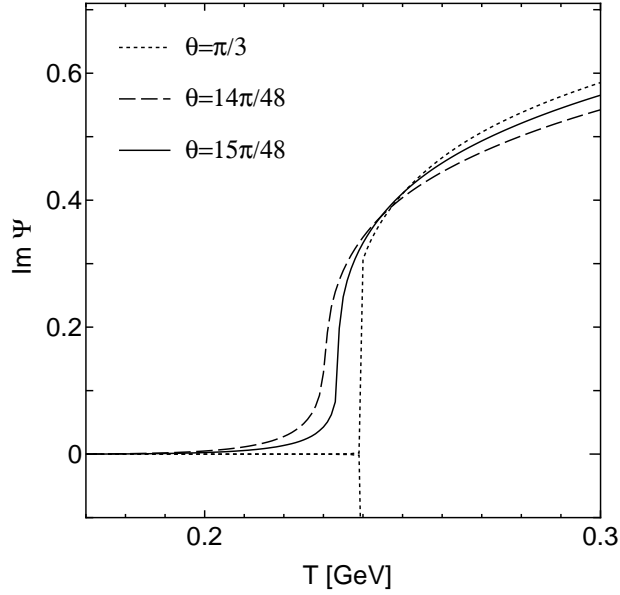}
\end{center}
\vspace{-.5cm}
\caption{ The $T$ dependence of $\mathrm{Im}~\Psi$ in the vicinity of the RW endpoint.
The dotted, dashed and solid lines are nonlocal PNJL results (using set I) at 
$\theta = \pi/3$ and its close neighborhood. 
}
\label{Fig:ImPsi_Tdep}
\end{figure}

Next we examine the region in the vicinity of the RW endpoint at $\theta=\pi/3$.
Figure~\ref{Fig:ImPsi_Tdep} shows the $T$ dependence of the imaginary part
$\mathrm{Im}\,\Psi$ of the modified Polyakov loop in the nonlocal PNJL model.
The dotted, dashed, and solid lines show results obtained at selected values $\theta=\pi/3$, $14\pi/48$, and $15\pi/48$, respectively, of the imaginary chemical potential divided by temperature.
One observes that $\mathrm{Im}~\Psi$, the order parameter of spontaneous $C$-symmetry breaking, develops a gap near $\theta = \pi/3$.
Other quantities such as $\sigma$ and $\mathrm{Re}~\Psi$ also display gaps at the same position, reflecting the coexistence condition for nonanalyticities of order parameters \cite{BCPG,Kashiwa3}.
However, this first-order behavior starts being suppressed as soon as $\theta$ deviates from $\theta=\pi/3$. This behavior is associated with explicit $C$-symmetry breaking becoming strong below the RW endpoint temperature $T_\mathrm{E}$.

\begin{figure}[htbp]
\begin{center}
 \includegraphics[width=0.45\textwidth]{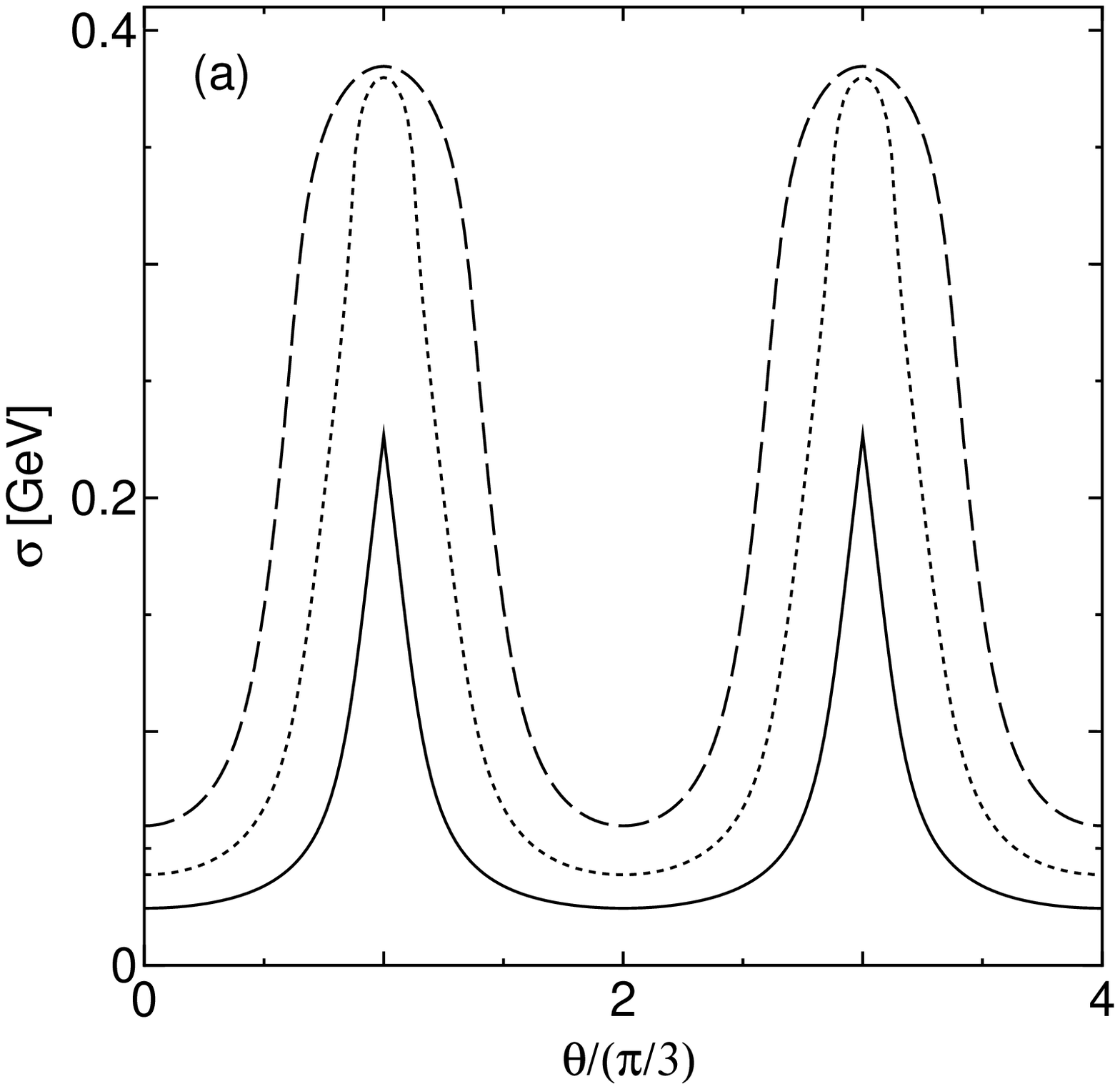}\\\vspace{0.3cm}
 \includegraphics[width=0.45\textwidth]{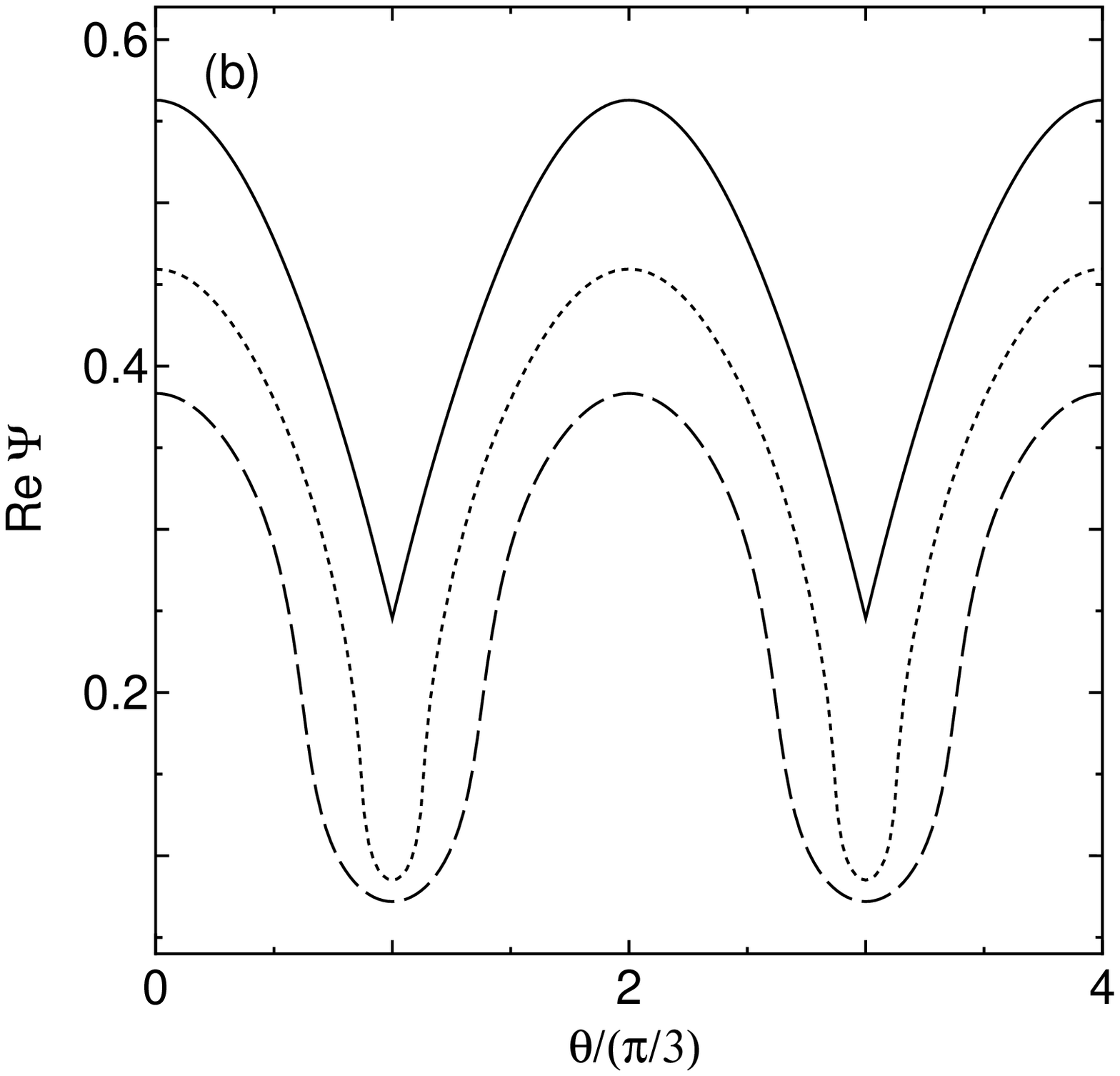}\hfill
 \includegraphics[width=0.45\textwidth]{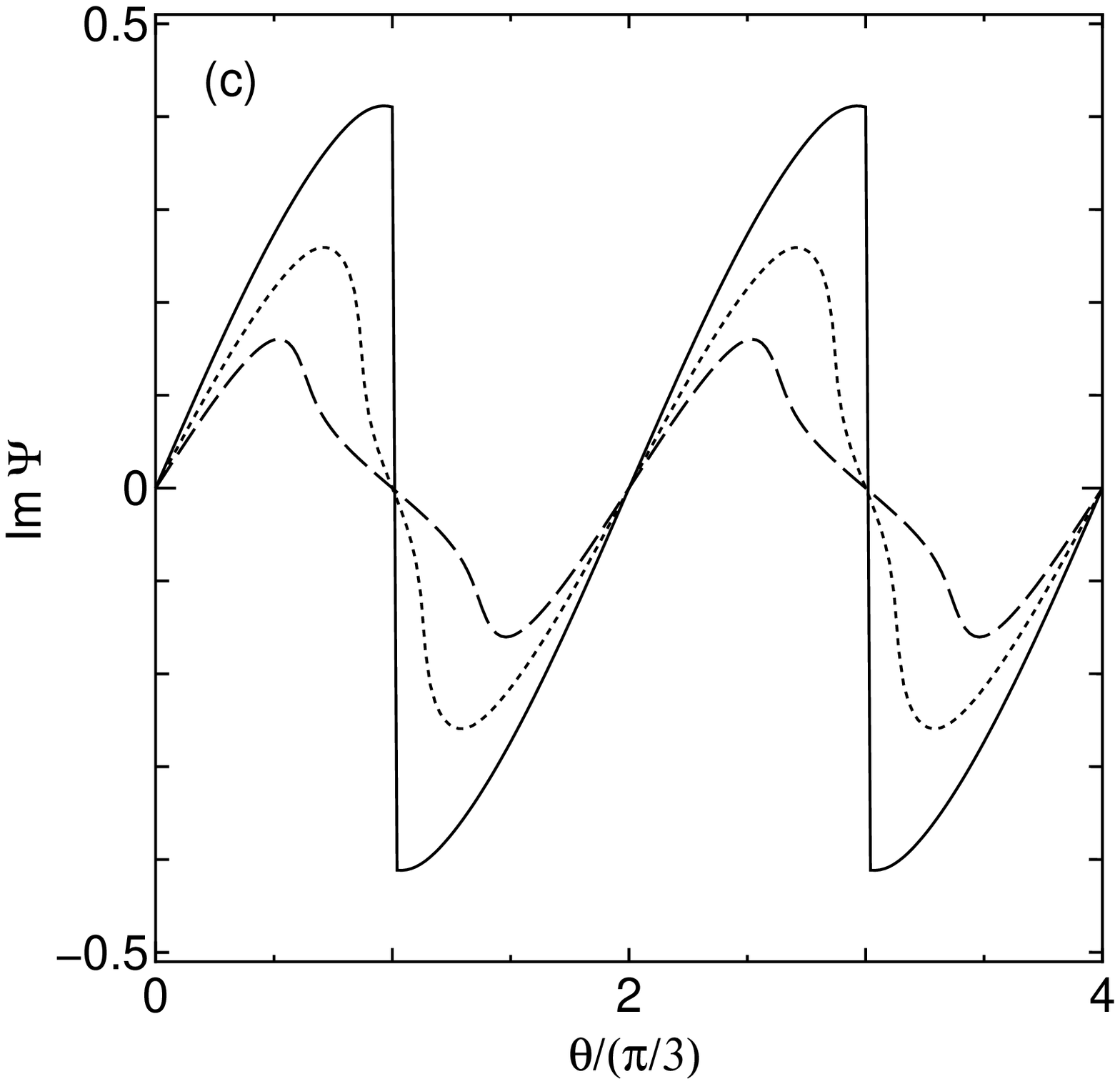}
\end{center}
\vspace{-.5cm}
\caption{ The $\theta$ dependences of the scalar mean field ${\sigma}$ [subfigure (a)], and of the modified Polyakov-loop, $\mathrm{Re}~\Psi$ and $\mathrm{Im}~\Psi$ (subfigures (b) and (c)).
The dashed, dotted and solid lines are results of nonlocal PNJL calculations (set I) at temperatures $T=220$, $230$ and $250\,\text{MeV}$, respectively.
}
\label{Fig:order_parameters_qdep}
\end{figure}

Figures~\ref{Fig:order_parameters_qdep} (a), (b) and (c) display the 
$\theta$ dependences of the order parameters ${\sigma}$, $\mathrm{Re}~\Psi$ and
$\mathrm{Im}~\Psi$, again calculated using the nonlocal PNJL model with set I.
The dashed, dotted and solid lines are the results at $T=220$, $230$ and $250\,\text{MeV}$, respectively.
At $T= 220$ and $230\,\text{MeV}$, these quantities show smooth behavior around $\theta=\pi /3$. At $T=250\,\text{MeV}$, on the other hand, the $\theta$-even functions ${\sigma}$ and $\mathrm{Re}~\Psi$ have sharp cusps at $\theta=\pi/3$. At the same temperature, the $\theta$-odd quantity $\mathrm{Im}~\Psi$ develops a characteristic gap at $\theta=\pi/3$. This is the RW transition mentioned earlier.
These results confirm that our nonlocal PNJL model is indeed capable of reproducing the RW periodicity as well as the RW transition in a way consistent with QCD.

It is important to note that, in the nonlocal PNJL model, the RW properties just described emerge from the consistent
insertion of the Polyakov-loop field in the distributions, ${\cal C}(p^2=\omega_{n,i}^2+\vec{p}\,^2)$ and ${\cal F}(p^2=\omega_{n,i}^2+\vec{p}\,^2)$, with $\omega_{n,i}$ given by Eq.~(\ref{eq17}). Had we omitted the combinations of gluonic background fields $A_4^3$ and $A_4^8$ in the shifted Matsubara frequencies $\omega_{n,i}$\,, the extended ${\mathbb Z}_3$ symmetry would have been explicitly broken, with the consequence of losing the RW periodicity and maintaining only the $2\pi$ periodicity in $\theta = \mu_\mathrm{I}/T$. The dependence of the chiral four-quark coupling on the $A_4^{3,8}$ fields generating the Polyakov loop, as discussed in Refs.~\cite{Kondo, Sakai3},  is thus a natural consequence of the RW periodicity requirement.  The nonlocal PNJL approach makes this important relationship explicit.

\begin{figure}[htbp]
\begin{center}
 \includegraphics[width=0.5\textwidth]{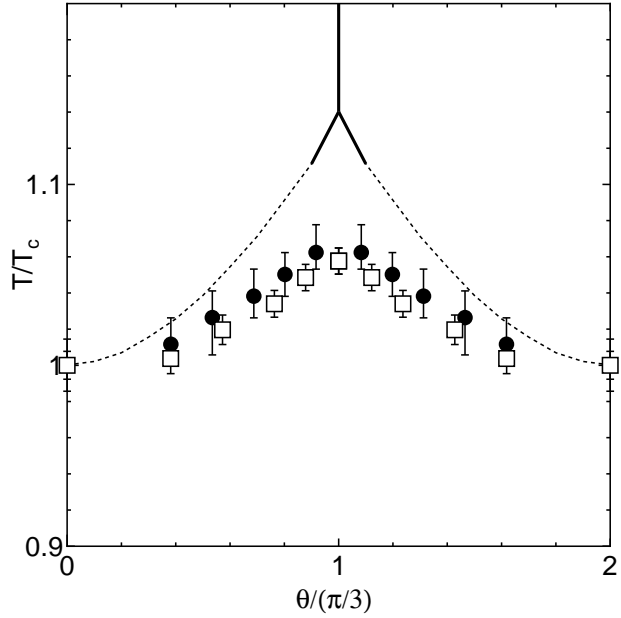}
\end{center}
\vspace{-.5cm}
\caption{ Phase diagram of the nonlocal PNJL model (set I) in the $T$-$\theta$ plane, in comparison with  LQCD results. 
The dotted and solid lines represent crossover and first-order transitions, respectively.
Open squares and closed circles are LQCD data taken from Refs.~\cite{Forcrand1} and Ref.~\cite{Wu}.
The temperature is given in units of the pseudo-critical temperature $T_c$ at $\mu = 0$, while the imaginary chemical potential is expressed as $\theta = \mu_\mathrm{I}/T$ in units of $\pi/3$.}
\label{Fig:PhaseDiagramI}
\end{figure}

Figure~\ref{Fig:PhaseDiagramI} shows the phase diagram at finite $\theta$ in the nonlocal PNJL model.
The solid lines represent first-order phase transitions while the dotted lines describe crossovers.
The symbols are LQCD data taken from Refs.~\cite{Wu, Forcrand1} [here we translate $\beta(a)/\beta(0)$ of the LQCD data to $T/T_c$ using the two-loop perturbative solution to the renormalization-group equation relating the lattice spacing $a$ and the lattice gauge coupling $\beta(a)$]. The RW endpoint becomes a triple point at which three first-order transition lines merge. In this nonlocal PNJL model the transition lines determined by ${\sigma}$ and $\mathrm{Im}~\Psi$ coincide in the entire $\theta$ region. This confirms once again that the two crossover lines to the left and right of $\theta = \pi/3$ are remnants of spontaneous $C$-symmetry breaking in this model.

\begin{figure}[htbp]
\begin{center}
 \includegraphics[width=0.5\textwidth]{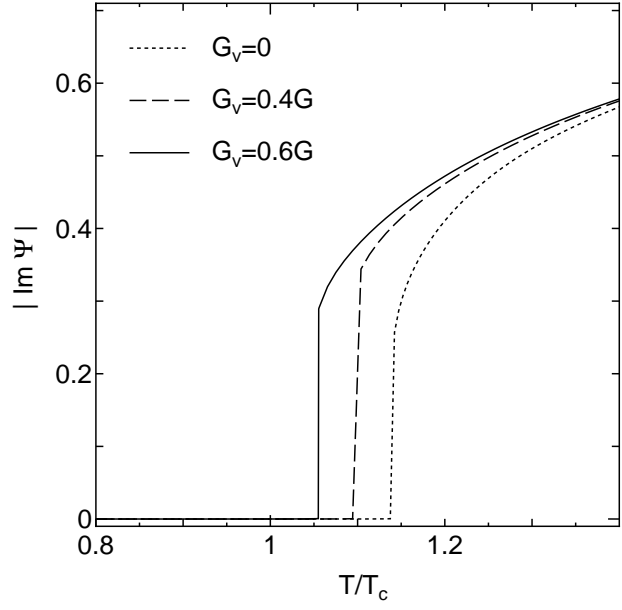}
\end{center}
\vspace{-.5cm}
\caption{ 
The $T$ dependence of ${\rm Im}~\Psi$ in the nonlocal PNJL + $G_\mathrm{v}$ model at $\theta=\pi/3$.
The dotted, dashed and solid lines are results obtained with $G_\mathrm{v}/G = 0$, $0.4$ and $0.6$, respectively.}
\label{Fig:G_v-dep}
\end{figure}

Next, we turn to set II and investigate the effects of the additional vector interaction, Eqs.\,(\ref{eq28}) and (\ref{vector}), as a function of the vector coupling strength $G_\mathrm{v}$. Figure \ref{Fig:G_v-dep} shows the $G_\mathrm{v}$ dependence of the order parameter ${\rm Im}\,\Psi$ at $\theta = \pi/3$. The primary effect of the additional vector-type four-quark interaction is evidently a reduction of the temperature at which the triple point appears. This shift is welcome recalling the comparison with LQCD data in 
Fig.~\ref{Fig:PhaseDiagramI}. At this stage a coupling $G_\mathrm{v}/G
\simeq 0.6$, slightly above the preferred range $G_\mathrm{v}/G \lesssim
0.5$ (see the Appendix), would give good agreement with lattice results.

\begin{figure}[htbp]
\begin{center}
 \includegraphics[width=0.5\textwidth]{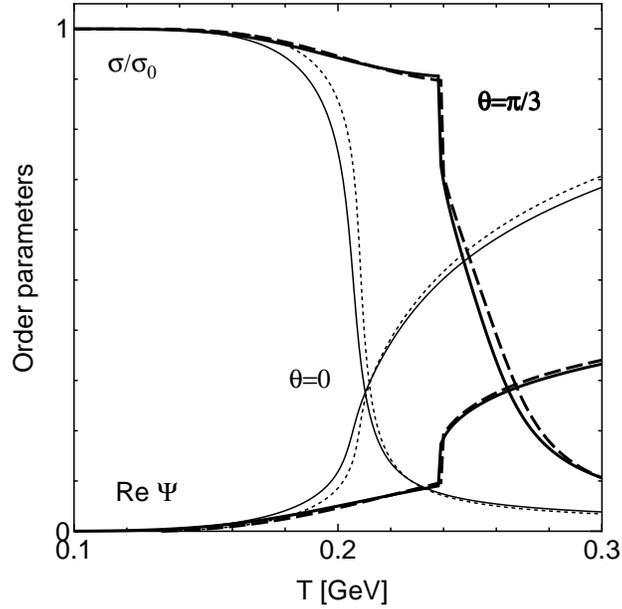}
\end{center}
\vspace{-.5cm}
\caption{The $T$ dependence of the order parameters $\sigma$ and $\mathrm{Re}\,\Psi$ at 
$\theta = 0$ and $\theta = \pi/3$ in the nonlocal  
PNJL model (set III) with inclusion of the quark quasiparticle wave-function-renormalization factor $Z(p^2)$ (solid curves) as compared to the case with $Z \equiv 1$
(dotted and dashed lines).}
\label{Fig:OP-Z2}
\end{figure}

In the next step we examine the influence of the quasiparticle $Z$ factor using the set III version of the nonlocal PNJL model.
Figure~\ref{Fig:OP-Z2} shows the $T$ dependence of the order parameters at $\theta=\pi/3$ and $\theta=0$, with and without wave function renormalization.
Evidently, the $Z$ factor does not affect the order parameters in any significant way over the entire range of $T$ and $\mu$. 

\begin{figure}[htbp]
\begin{center}
 \includegraphics[width=0.5\textwidth]{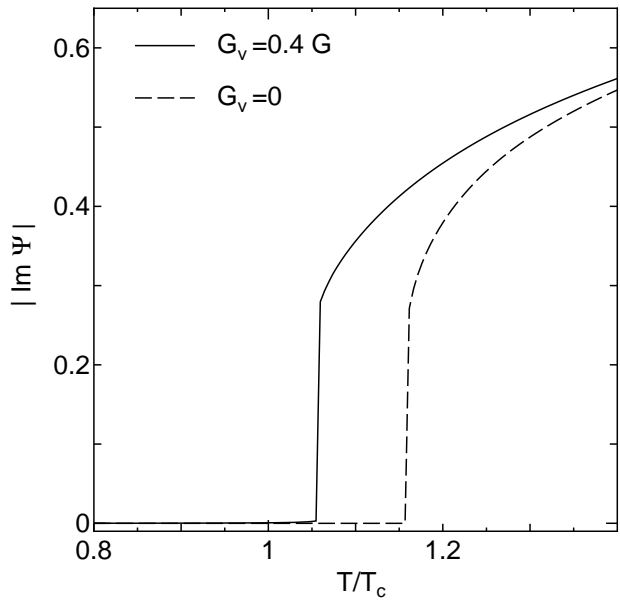}
\end{center}
\vspace{-.5cm}
\caption{ The $G_\mathrm{v}$ dependence of $\mathrm{Im}~\Psi$ in the nonlocal PNJL model (set IV).
The dashed and solid lines denote the results at $\theta=\pi/3$ with $G_\mathrm{v}=0$ and 
$G_\mathrm{v}=0.4\,G$, respectively.
}
\label{Fig:Fig-ImPsi-FF-ZZ}
\end{figure}

\begin{figure}[htbp]
\begin{center}
 \includegraphics[width=0.5\textwidth]{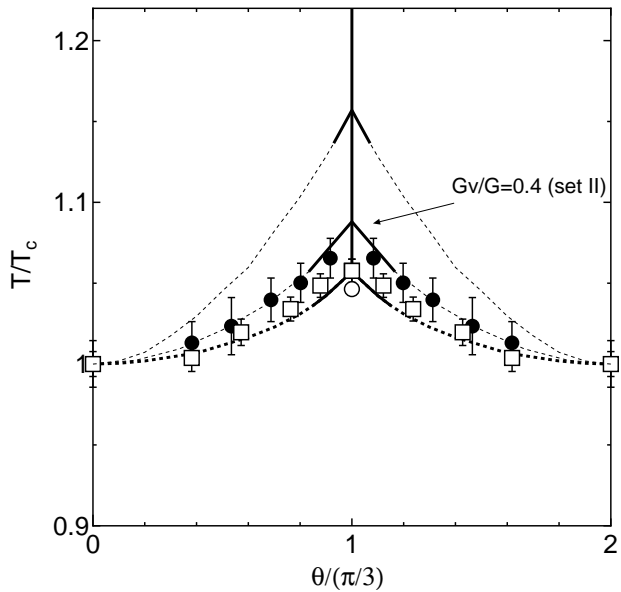}
\end{center}
\vspace{-.5cm}
\caption{ Phase diagram of
the nonlocal PNJL model with inclusion of the quark quasiparticle renormalization factor $Z(p^2)$. 
Upper curves: without additional vector interaction (model set III). Lower curves: including vector coupling with $G_\mathrm{v}/G = 0.4$ (model set IV). 
The dotted and solid lines represent crossover and first-order phase transitions, respectively. Also shown for comparison is the result with $G_\mathrm{v}/G = 0.4$ but with no Z-factor effect (set II).
Open squares and closed circles are LQCD data taken from Ref~\cite{Forcrand1} and Ref.~\cite{Wu}. The temperature scale $T_c$ is the pseudo-critical temperature at $\mu=0$.
The open circle at the triple point is the result of the set IV calculation with quark mass $m_0=12\,\text{MeV}$.
}
\label{Fig:PhaseDiagramIZ}
\end{figure}

Finally, moving to set IV the combined effects of the additional vector interaction and of the $Z$ factor are studied. Figure~\ref{Fig:Fig-ImPsi-FF-ZZ} shows the dependence of $\mathrm{Im}~\Psi$ on the vector coupling strength $G_\mathrm{v}$. 
The additional role of the $Z$ factor is now to enhance the vector coupling effect in the downward shift of the temperature at which the  triple point appears. In fact, good agreement with LQCD results is now achieved using $G_\mathrm{v}/G = 0.4$, well within the natural range of  $G_\mathrm{v}$ implied by the Fierz-transformed color-current-current interaction between quarks. The resulting phase diagram is displayed in  Fig.~\ref{Fig:PhaseDiagramIZ} along with LQCD data. A direct comparison must take into account the fact that these data have been generated with a quark mass $m_0 = 12\,\text{MeV}$, about $4$ times the standard input current quark mass in the PNJL calculation. The pion mass corresponding to this heavier quark mass is thus twice as large as the physical one. We have studied the quark mass dependence in the vicinity of the triple point. 
The open circle in Fig.\,\ref{Fig:PhaseDiagramIZ} is the PNJL result using $m=12\,\text{MeV}$. Evidently, the difference in 
$T/T_c$ between calculations with physical pion mass and with $m_\pi\simeq280\,\text{MeV}$ is only marginal.

\begin{figure}[htbp]
\begin{center}
 \includegraphics[width=0.5\textwidth]{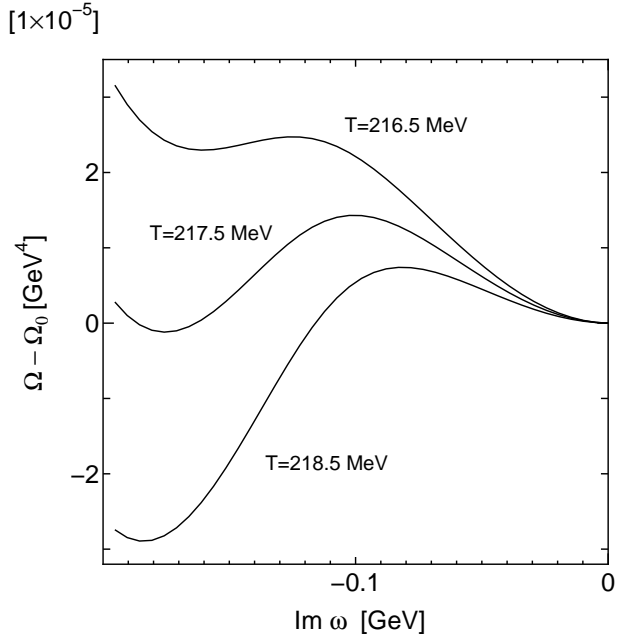}
\end{center}
\vspace{-.5cm}
\caption{ 
Thermodynamical potential $\Omega$ of the nonlocal PNJL model (set IV with $G_\mathrm{v} = 0.4\, G$) as a function of the imaginary vector mean field ${\rm Im}~\omega$.
The lines are results of calculations at neighboring temperatures $T=216.5$, $T=217.5$ and $218.5\,\text{MeV}$, as indicated. 
The reference $\Omega_0$ is defined at ${\rm Im}~\omega=0$.}
\label{Fig:Omega}
\end{figure}

The behavior of the thermodynamic potential $\Omega$ as a function of the (imaginary) isoscalar-vector mean field, 
${\rm Im}~\omega$, is also of some interest here in order to examine details of the phase transition. 
Figure\,\ref{Fig:Omega} shows $\Omega$ for set IV, once again using $G_\mathrm{v}/G = 0.4$, at three neighboring temperatures, demonstrating the occurrence of a first-order transition near $T=217.5\,\text{MeV}$.  

\section{Summary}

In this paper we have investigated properties of two-flavor QCD at finite imaginary quark chemical potential
$\mu_\mathrm{I}$ in order to establish constraints for the modeling of the QCD phase diagram. A recently extended version of the nonlocal PNJL model has been used for this task. Apart from the momentum-dependent quasiparticle mass $M(p^2)$ of the quarks, and their gauge-covariant coupling to a Polyakov-loop background field, this model also incorporates the quark wave-function-renormalization factor, $Z(p^2)$. Both $M(p^2)$ and $Z(p^2)$ are introduced in close correspondence with results from Dyson-Schwinger calculations and LQCD computations. We can draw the following conclusions:

1. The nonlocal PNJL model is capable of reproducing important generic properties of QCD at $\mu_\mathrm{I} > 0$: the Roberge-Weiss (RW) periodicity and the RW phase transition. It turns out that the RW endpoint located at odd multiples of $\theta = \mu_\mathrm{I} / T = \pi/3$ is first order. It is actually a triple point at which three first-order phase transition lines merge, signaling spontaneous breaking of charge-conjugation symmetry. This first-order transition region in the $T$-$\theta$ phase diagram is restricted to a small vicinity of the triple point, outside of which explicit $C$-symmetry breaking takes over and turns the transitions into continuous crossover lines extending symmetrically left and right of the triple-point area. 

2. The (pseudo-)critical temperatures for the chiral and deconfinement crossovers and first-order transitions turn out to coincide both at zero chemical potential and at $\mu_\mathrm{I} > 0$. This coincidence is a characteristic feature of the {\it nonlocal} PNJL model. It is based on the consistent treatment of the Polyakov-loop dependence of the nonlocality distributions that govern the effective interactions between quark quasiparticles.

3. Isoscalar-vector interactions between quarks, additional to the standard chiral combination of isovector-pseudoscalar and isoscalar-scalar interactions, play an important role in the $T$-$\mu$ phase diagram at imaginary $\mu$. Such vector interactions are well known to emerge naturally from Fierz-transforming the basic color-current-current interaction of QCD. Their presence is found necessary in order to reproduce the location of the
RW endpoint on the $T$ scale and the pattern of transition lines in the neighborhood of that point as observed in LQCD. 

4. The wave-function-renormalization factor $Z(p^2)$, resulting from derivative couplings between quarks, does not play an important role individually. However, in combination with the isoscalar-vector interaction just mentioned, it has a visible impact on the phase diagram. The $Z$ factor and the vector coupling cooperate coherently in reproducing the phase diagram at $\mu_\mathrm{I} > 0$ from LQCD. Including both effects the resulting vector coupling strength, relative to that of the scalar interaction that generates spontaneous chiral symmetry breaking, is found to be 
$G_\mathrm{v} = 0.4\,G$. This value is remarkably consistent with earlier studies of meson properties and spectroscopy within NJL models including realistic constraints from the QCD axial anomaly.

\section{Appendix:\\
Color-current-current correlations and isoscalar-vector interaction}

Here we first sketch the derivation (see Ref.~\cite{Kondo}) of an effective color-current-current interaction from two-flavor QCD and then recall its Fierz transform leading to NJL-type four-point couplings (see e.g. Refs.~\cite{Vogl,Hatsuda}).
Consider the QCD Lagrangian 
\begin{align}
{\cal L}_\mathrm{QCD} &= {\cal L}_{q} + {\cal L}_\mathrm{YM} + {\cal L}_{qG}~,\\
{\cal L}_{q}  &= {\bar q}(x) (i \gamma_\mu \partial^\mu -m ) q(x)~,\\
{\cal L}_\mathrm{YM}  &= -\frac{1}{4}F_{\mu\nu}^a(x) F^{\mu\nu}_a(x)~, \\
{\cal L}_{qG} &= g\, {\bar q}(x)\, \gamma_\mu A^{\mu}(x)\, q(x)~,
\end{align}
with $q(x)=(u(x),d(x))^T$ and  $A^{\mu} \equiv A^a_\mu \,\lambda_a /2$.
The QCD generating functional becomes
\begin{align}
\mathcal{Z}_\mathrm{QCD} 
&= \int {\cal D}q {\cal D} {\bar q}
   \int {\cal D} A^a_\mu 
   \exp \Bigl(
               i \int \mathrm{d}^4 x~{\cal L}_\mathrm{QCD}
         \Bigr) \nonumber\\
&= \int {\cal D}q {\cal D} {\bar q}
   \exp \Bigl[ i \int \mathrm{d}^4 x ({\cal L}_{q} + iW[j] ) \Bigr]~,
\end{align}
where 
\begin{align}
iW[j] &= \ln \int {\cal D} A_\mu^a 
         \exp \Bigl[ \int \mathrm{d}^4 x~ ({\cal L}_\mathrm{YM} + {\cal L}_{qG}) \Bigr]~,
\end{align}
involves the quark color currents
\begin{align}
j^a_\mu(x) &= {\bar q}(x)\, \frac{\lambda^a}{2} \gamma_\mu\, q(x)~.
\end{align}
Expanding the generating functional in powers of this color current, the function $iW$ can be expressed as
\begin{align}
iW[j] &= iW[0] - g \int \mathrm{d}^4x\,W^{(1)a}_{\mu}(x)\, j_{a}^\mu(x)
\nonumber\\
&              + \frac{g^2}{2} 
                 \int d^4x \int \mathrm{d}^4y\,  j_{a}^\mu(x)\,W^{(2)ab}_{\mu\nu}(x,y)\, j_{b}^\nu(y)
               + {\cal O}(j^3)~,
\end{align}
where $W^{(n)}$ is related to the connected gluon $n$-point function without quark loops. 
Truncating this expansion at $W^{(2)}$, this becomes a model in which the {\it local} color gauge symmetry of QCD is reduced to a {\it global} $\mathrm{SU}(3)_c$ symmetry [global color model (GCM)], with the reduced Lagrangian 
\begin{align}
{\cal L}_\mathrm{GCM}(x) &= {\bar q}(x)\, (i \gamma_\mu \partial^\mu - m )\, q(x)  
                + g^2 j_{a}^{\mu}(x) \int \mathrm{d}^4y\,W^{(2)ab}_{\mu\nu}(x,y)
                                      \,j_{b}^{\nu}(y) ~.
\end{align}
This Lagrangian features the nonlocal color-current interaction that involves the full gluon propagator.
The nonlocal PNJL approach basically keeps this structure. Assume now that the gluonic correlator determining $W^{(2)}$ is very short-ranged so that it can be localized with a leading term proportional to $\delta^4(x-y)$. This is the idea behind the classic NJL and local PNJL models. With this local approximation the interaction Lagrangian becomes
\begin{align}
{\cal L}_\mathrm{int}(x) &= G_c \sum_{a=1}^8\, 
                  [{\bar q}(x)\, t^a \gamma_\mu \,q(x)]\,
                  [{\bar q}(x)\, t^a \gamma^\mu \,q(x)] \nonumber\\
&= G_c \sum_{a=1}^8 
   {\bar q}_{r \alpha}^i (\,t^a)_{\alpha \beta} \,
      (\gamma_\mu)_{rs}\, \delta^{ij} q^j_{s\beta}\,
    {\bar q}_{t \gamma}^k \,(t^a)_{\gamma \delta} \,
      (\gamma^\mu)_{tu}\, \delta^{kl} \,q^l_{u \delta}~,
\label{Eq:color-current-l}
\end{align}
with the $\mathrm{SU}(3)_c$ generators $t^a = \lambda^a/2$. Here $G_c$ is an effective color coupling strength of dimension $[\text{length}]^2$. The second part of Eq.~(\ref{Eq:color-current-l}) explicitly displays the combinations and contractions of quark color indices $\alpha, \beta, \ldots\in \{1,2,3\}$, flavor (isospin) indices $i, j, \ldots \in \{1,2\}$ and Dirac spinor indices $r, s, \ldots\in \{1,\ldots, 4\}$.
This interaction represents color-octet exchange between quarks. A Fierz
transform translates this into various color-singlet channels and color-octet quark-antiquark channels with isospin singlet or triplet quantum numbers.

Consider first the general $\mathrm{SU}(N)$ Fierz transformation from channels $(12\rightarrow 34)$ to the exchange channels $(14\rightarrow 32)$:
\begin{align}
( {\bf 1} \otimes {\bf 1} )_{12,34} 
&= \frac{1}{N} ( {\bf 1} \otimes {\bf 1} )_{14,32}  + 2 ( t^a \otimes t^a)_{14,32}~, 
\label{Firz1}
\\
(t^a \otimes t^a)_{12,34} 
&= \frac{1}{2} \left( 1 - \frac{1}{N^2} \right) ( {\bf 1} \otimes {\bf 1})_{14,32} 
 - \frac{1}{N} ( t^a \otimes t^a )_{14,32}~.
\label{Firz2}
\end{align}
The $``\bf 1"$ symbolizes the $N\times N$ unit matrix while the matrices $t^a$ stand for the generators $\tau^a/2$ or $\lambda^a/2$ of $\mathrm{SU}(2)$ or $\mathrm{SU}(3)$, respectively. Using this expression, the Fierz transform of the color degrees of freedom
in Eq.~(\ref{Eq:color-current-l}) becomes 
\begin{align}
(t^a \otimes t^a)_{\alpha \beta, \gamma \delta} 
&= \frac{1}{2} \left( 1-\frac{1}{N_c^2} \right) 
  ( {\bf 1} \otimes {\bf 1} )_{\alpha \delta, \gamma \beta}
 - \frac{1}{N_c} ( t^a \otimes t^a)_{\alpha \delta, \gamma \beta}~, 
\end{align}
where the first term denotes the color-singlet and the second one refers to the color-octet channels.
In the present study we consider only the color-singlet part and absorb the coefficient $4/9$ multiplying $( {\bf 1} \otimes {\bf 1} )$ in a redefined coupling constant $G_c$.

Next, consider the Fierz transformation in Dirac space. The well-known result is
\begin{align}
(\gamma_\mu \otimes \gamma^\mu)_{rs,tu}
&= - ({\bf 1} \otimes {\bf 1} + i\gamma_5 \otimes i \gamma_5)_{ru,ts}
   + \frac{1}{2} ( \gamma_\mu \otimes \gamma^\mu 
                 + \gamma_\mu \gamma_5 \otimes \gamma^\mu \gamma_5)_{ru,ts}~.
\end{align}
Finally, the Fierz transformation in flavor-$\mathrm{SU}(2)$ space, following Eq.~(\ref{Firz1}), gives in terms of the isospin Pauli matrices $\tau^i$:
\begin{align}
({\bf 1} \otimes {\bf 1})_{ij,kl}
&= \frac{1}{2} ({\bf 1} \otimes {\bf 1})_{il,kj} 
 + \frac{1}{2} \sum_{i=1}^3 (\tau^i \otimes \tau^i)_{il,kj}~.
\end{align}
The color-singlet-interaction Lagrangian derived from the color-current-current interaction becomes
\begin{align}
{\cal L}_\mathrm{int}^{(0)} 
&= {G\over 2} \Bigl[ 
       ({\bar q}q)^2 + ({\bar q}\, i \gamma_5\, q)^2 
     + ({\bar q}\, {\vec \tau}\, q)^2 + ({\bar q}\, i \gamma_5 \,{\vec \tau} \,q)^2
\nonumber\\
& \hspace{0.6cm}   
     - \frac{1}{2}
       \Bigl\{
       ({\bar q} \,\gamma^\mu \,q)^2 + ({\bar q} \,\gamma_\mu \gamma_5 \,q)^2 
     + ({\bar q}\, \gamma^\mu\, {\vec \tau}\, q)^2 
     + ({\bar q}\, \gamma_\mu \gamma_5 \,{\vec \tau}\, q)^2 
       \Bigr\}
      \Bigr]~.
\end{align}
This interaction still has a chiral $\mathrm{U}(2)_\mathrm{R} \times \mathrm{U}(2)_\mathrm{L}$ symmetry.
An axial $\mathrm{U}(1)$-breaking term must be added in order to account for the axial anomaly in QCD and reduce
the symmetry to chiral $\mathrm{SU}(2)_\mathrm{R} \times \mathrm{SU}(2)_\mathrm{L}$, times $\mathrm{U}(1)_\mathrm{V}$ for conserved baryon number. 
The $\mathrm{U}(1)_\mathrm{A}$ anomaly effect is introduced by the Kobayashi-Maskawa-'t Hooft determinant interaction as
\begin{align}
{\cal L}_\mathrm{anomaly}
&= G_A \det [{\bar q}\,(1+\gamma_5)\, q ] + h.c. 
\nonumber\\
&= {G_A\over 2}
       \Bigl[ ({\bar q}q)^2 + ({\bar q}\, i\gamma_5\, {\vec \tau}\, q)^2 
            - ({\bar q}\, {\vec \tau}\, q)^2 - ({\bar q}\, i \gamma_5\, q )^2
       \Bigr]~,
\label{eq:minimal}
\end{align}
where $\det$ acts in flavor space.
Setting $G_A=G$ leads to the usual NJL-type Lagrangian (see Refs.~\cite{Klevansky,Vogl,Hatsuda,Buballa} for reviews of the NJL model). In this case the total interaction becomes
\begin{align}
{\cal L}_\mathrm{int} 
= {\cal L}_\mathrm{int}^{(0)} + {\cal L}_\mathrm{anomaly} &=  G \Bigl[ 
       ({\bar q}q)^2 + ({\bar q}\, i \gamma_5 \vec{\tau}q)^2 
\nonumber\\
 \hspace{0.6cm}   
     &- \frac{1}{4}
       \Bigl\{
       ({\bar q} \gamma^\mu q)^2 + ({\bar q} \gamma_\mu \gamma_5 q)^2 
     + ({\bar q} \gamma^\mu {\vec \tau} q)^2 
     + ({\bar q} \gamma_\mu \gamma_5 {\vec \tau} q)^2 
       \Bigr\}
      \Bigr]~.
\label{eq:maximal}
\end{align}
This is the situation with ``maximal" $\mathrm{U}(1)_\mathrm{A}$ symmetry breaking in which the anomaly term is strong and accounts for half of the scalar-isoscalar and pseudoscalar-isovector interactions, while the scalar-isovector and pseudoscalar-isoscalar terms are completely eliminated. In this case the vector coupling strength is $G_\mathrm{v} = 0.25\,G$, denoting the vector interaction term as $-G_\mathrm{v}\,(\bar{q}\gamma_\mu q)^2$. The instanton liquid model \cite{Schaufer,Rapp} produces such a scenario. However, the strength of the anomaly-driven part of the interaction is subject to ambiguity. 
In Refs.~\cite{Klimt,Frank} the percentage for the anomaly term in the total scalar-isoscalar and pseudoscalar-isovector coupling constant was analyzed using the three-flavor NJL model. According to these evaluations, the anomaly term accounts for about one fifth (more precisely: 16\,\%--21\,\%) of the total effective four-quark interaction with parameters as given in Refs.~\cite{Kunihiro,Klimt,Rehberg}. In this case the resulting ratio of vector-to-scalar couplings is $G_\mathrm{v}/G = 0.4$. In practice the acceptable range for this ratio is taken to be $0.25\le G_\mathrm{v}/G \lesssim 0.5$ with a preference for values around 0.4. 


\noindent
\begin{acknowledgments}
One of the authors (K.K.) thanks H. Kouno and T. Matsumoto for useful discussions.
He acknowledges support by the Japan Society for the Promotion of Science for Young Scientists and RIKEN.
This work is supported in part by BMBF and by the DFG Excellence Cluster 
``Origin and Structure of the Universe''.
\end{acknowledgments}



\begin{thebibliography}{19}
\expandafter\ifx\csname natexlab\endcsname\relax\def\natexlab#1{#1}\fi
\expandafter\ifx\csname bibnamefont\endcsname\relax
  \def\bibnamefont#1{#1}\fi
\expandafter\ifx\csname bibfnamefont\endcsname\relax
  \def\bibfnamefont#1{#1}\fi
\expandafter\ifx\csname citenamefont\endcsname\relax
  \def\citenamefont#1{#1}\fi
\expandafter\ifx\csname url\endcsname\relax
  \def\url#1{\texttt{#1}}\fi
\expandafter\ifx\csname urlprefix\endcsname\relax\def\urlprefix{URL }\fi
\providecommand{\bibinfo}[2]{#2}
\providecommand{\eprint}[2][]{\url{#2}}
%


\bibitem[{\citenamefont{Meisinger}(1996)}]{Meisinger}
\bibinfo{author}{\bibfnamefont{P.}~\bibfnamefont{N.}~\bibnamefont{Meisinger}}, 
\bibnamefont{and}
\bibinfo{author}{\bibfnamefont{M.}~\bibfnamefont{C.}~\bibnamefont{Ogilvie}},  
  \bibinfo{journal}{Phys. Lett.\ B} \textbf{\bibinfo{volume}{379}},
  \bibinfo{pages}{163} (\bibinfo{year}{1996}).

\bibitem[{\citenamefont{Fukushima}(2004)}]{Fukushima1}
\bibinfo{author}{\bibfnamefont{K.}~\bibnamefont{Fukushima}}, 
  \bibinfo{journal}{Phys. Lett.\ B} \textbf{\bibinfo{volume}{591}},
  \bibinfo{pages}{277} (\bibinfo{year}{2004});
  \bibinfo{journal}{Phys. Rev. D} \textbf{\bibinfo{volume}{77}},
  \bibinfo{pages}{114028} (\bibinfo{year}{2008});
  \bibinfo{journal}{Phys. Rev. D} \textbf{\bibinfo{volume}{78}},
  \bibinfo{pages}{114019} (\bibinfo{year}{2008}).

\bibitem[{\citenamefont{Ratti et al.}(2006)}]{Ratti}
\bibinfo{author}{\bibfnamefont{C.}~\bibnamefont{Ratti}},
\bibinfo{author}{\bibfnamefont{M.}~\bibfnamefont{A.}~\bibnamefont{Thaler}},
\bibnamefont{and}
\bibinfo{author}{\bibfnamefont{W.}~\bibnamefont{Weise}},  
  \bibinfo{journal}{Phys. Rev.\ D} \textbf{\bibinfo{volume}{73}},
  \bibinfo{pages}{014019} (\bibinfo{year}{2006}). 

\bibitem[{\citenamefont{Ciminale et al.}(2008)}]{Ciminale}
\bibinfo{author}{\bibfnamefont{M.}~\bibnamefont{Ciminale}},
\bibinfo{author}{\bibfnamefont{R.}~\bibnamefont{Gatto}},
\bibinfo{author}{\bibfnamefont{N.}~\bibfnamefont{D.}~\bibnamefont{Ippolito}},
\bibinfo{author}{\bibfnamefont{G.}~\bibnamefont{Nardulli}},
\bibnamefont{and}
\bibinfo{author}{\bibfnamefont{M.}~\bibnamefont{Ruggieri}},  
  \bibinfo{journal}{Phys. Rev.\ D} \textbf{\bibinfo{volume}{77}},
  \bibinfo{pages}{054023} (\bibinfo{year}{2008}). 

\bibitem[{\citenamefont{Kashiwa et al}(2008)}]{Kashiwa1}
\bibinfo{author}{\bibfnamefont{K.}~\bibnamefont{Kashiwa}}, 
\bibinfo{author}{\bibfnamefont{H.}~\bibnamefont{Kouno}}, 
\bibinfo{author}{\bibfnamefont{M.}~\bibnamefont{Matsuzaki}}, 
\bibnamefont{and}
\bibinfo{author}{\bibfnamefont{M.}~\bibnamefont{Yahiro}},
  \bibinfo{journal}{Phys.\ Lett.\  B} \textbf{\bibinfo{volume}{662}},
  \bibinfo{pages}{26} (\bibinfo{year}{2008}).

\bibitem[{\citenamefont{Kashiwa et al}(2008)}]{Abuki}
\bibinfo{author}{\bibfnamefont{H.}~\bibnamefont{Abuki}}, 
\bibinfo{author}{\bibfnamefont{R.}~\bibnamefont{Anglani}}, 
\bibinfo{author}{\bibfnamefont{R.}~\bibnamefont{Gatto}}, 
\bibinfo{author}{\bibfnamefont{G.}~\bibnamefont{Nardulli}}, 
\bibnamefont{and}
\bibinfo{author}{\bibfnamefont{M.}~\bibnamefont{Ruggieri}},
  \bibinfo{journal}{Phys. Rev. D} \textbf{\bibinfo{volume}{78}},
  \bibinfo{pages}{034034} (\bibinfo{year}{2008}).

\bibitem[{\citenamefont{Sakai et al}(2008)}]{Sakai1}
\bibinfo{author}{\bibfnamefont{Y.}~\bibnamefont{Sakai}},
\bibinfo{author}{\bibfnamefont{K.}~\bibnamefont{Kashiwa}}, 
\bibinfo{author}{\bibfnamefont{H.}~\bibnamefont{Kouno}}, 
\bibnamefont{and}
\bibinfo{author}{\bibfnamefont{M.}~\bibnamefont{Yahiro}},
  \bibinfo{journal}{Phys.\ Rev.\  D} \textbf{\bibinfo{volume}{77}},
  \bibinfo{pages}{051901(R)} (\bibinfo{year}{2008});
%
  \bibinfo{journal}{Phys.\ Rev.\  D} \textbf{\bibinfo{volume}{78}},
  \bibinfo{pages}{ 036001} (\bibinfo{year}{2008}); 
\bibinfo{author}{\bibfnamefont{Y.}~\bibnamefont{Sakai}},
\bibinfo{author}{\bibfnamefont{K.}~\bibnamefont{Kashiwa}}, 
\bibinfo{author}{\bibfnamefont{H.}~\bibnamefont{Kouno}}, 
\bibinfo{author}{\bibfnamefont{M.}~\bibnamefont{Matsuzaki}}, 
\bibnamefont{and}
\bibinfo{author}{\bibfnamefont{M.}~\bibnamefont{Yahiro}},
  \bibinfo{journal}{Phys.\ Rev.\  D} \textbf{\bibinfo{volume}{78}},
  \bibinfo{pages}{076007} (\bibinfo{year}{2008}).

\bibitem[{\citenamefont{Hell et al.}(2009)}]{Hell1}
\bibinfo{author}{\bibfnamefont{T.}~\bibnamefont{Hell}}, 
\bibinfo{author}{\bibfnamefont{S.}~\bibnamefont{R\"{o}{\ss}ner}}, 
\bibinfo{author}{\bibfnamefont{M.}~\bibnamefont{Cristoforetti}}, 
\bibnamefont{and}
\bibinfo{author}{\bibfnamefont{W.}~\bibnamefont{Weise}},  
\bibinfo{journal}{Phys. Rev. D} \textbf{\bibinfo{volume}{79}},
\bibinfo{pages}{014022} (\bibinfo{year}{2009}). 

\bibitem[{\citenamefont{Hell et al.}(2010)}]{Hell2}
\bibinfo{author}{\bibfnamefont{T.}~\bibnamefont{Hell}}, 
\bibinfo{author}{\bibfnamefont{S.}~\bibnamefont{R\"{o}{\ss}ner}}, 
\bibinfo{author}{\bibfnamefont{M.}~\bibnamefont{Cristoforetti}}, 
\bibnamefont{and}
\bibinfo{author}{\bibfnamefont{W.}~\bibnamefont{Weise}},  
\bibinfo{journal}{Phys. Rev. D} \textbf{\bibinfo{volume}{81}},
\bibinfo{pages}{074034} (\bibinfo{year}{2010}). 

\bibitem[{\citenamefont{Kouno et al}(2009)}]{Kouno}
\bibinfo{author}{\bibfnamefont{H.}~\bibnamefont{Kouno}},
\bibinfo{author}{\bibfnamefont{Y.}~\bibnamefont{Sakai}}, 
\bibinfo{author}{\bibfnamefont{K.}~\bibnamefont{Kashiwa}}, 
\bibnamefont{and}
\bibinfo{author}{\bibfnamefont{M.}~\bibnamefont{Yahiro}},
  \bibinfo{journal}{J. Phys. \  G: Nucl. Part. Phys.} \textbf{\bibinfo{volume}{36}},
  \bibinfo{pages}{115010} (\bibinfo{year}{2009}). 

\bibitem[{\citenamefont{Sakai et al}(2008)}]{Sakai2}
\bibinfo{author}{\bibfnamefont{Y.}~\bibnamefont{Sakai}},
\bibinfo{author}{\bibfnamefont{K.}~\bibnamefont{Kashiwa}}, 
\bibinfo{author}{\bibfnamefont{H.}~\bibnamefont{Kouno}}, 
\bibinfo{author}{\bibfnamefont{M.}~\bibnamefont{Matsuzaki}}, 
\bibnamefont{and}
\bibinfo{author}{\bibfnamefont{M.}~\bibnamefont{Yahiro}},
  \bibinfo{journal}{Phys.\ Rev.\  D} \textbf{\bibinfo{volume}{79}},
  \bibinfo{pages}{ 096001} (\bibinfo{year}{2009}). 

\bibitem[{\citenamefont{Kashiwa et al}(2009)}]{Kashiwa2}
\bibinfo{author}{\bibfnamefont{K.}~\bibnamefont{Kashiwa}}, 
\bibinfo{author}{\bibfnamefont{M.}~\bibnamefont{Matsuzaki}}, 
\bibinfo{author}{\bibfnamefont{H.}~\bibnamefont{Kouno}}, 
\bibinfo{author}{\bibfnamefont{Y.}~\bibnamefont{Sakai}},
\bibnamefont{and}
\bibinfo{author}{\bibfnamefont{M.}~\bibnamefont{Yahiro}},
  \bibinfo{journal}{Phys.\ Rev.\  D} \textbf{\bibinfo{volume}{79}},
  \bibinfo{pages}{076008} (\bibinfo{year}{2009}).

\bibitem[{\citenamefont{Kashiwa et al}(2009)}]{Kashiwa3}
\bibinfo{author}{\bibfnamefont{K.}~\bibnamefont{Kashiwa}}, 
\bibinfo{author}{\bibfnamefont{M.}~\bibnamefont{Yahiro}},
\bibinfo{author}{\bibfnamefont{H.}~\bibnamefont{Kouno}},
\bibinfo{author}{\bibfnamefont{M.}~\bibnamefont{Matsuzaki}}, 
\bibnamefont{and}
  \bibinfo{author}{\bibfnamefont{Y.}~\bibnamefont{Sakai}}, 
  \bibinfo{journal}{J. Phys. \  G: Nucl. Part. Phys.} \textbf{\bibinfo{volume}{36}},
  \bibinfo{pages}{105001} (\bibinfo{year}{2009}). 

\bibitem[{\citenamefont{Roberge and Weiss}(1986)}]{RW}
\bibinfo{author}{\bibfnamefont{A.}~\bibnamefont{Roberge}} 
\bibnamefont{and}
\bibinfo{author}{\bibfnamefont{N.}~\bibnamefont{Weiss}},  
\bibinfo{journal}{Nucl. Phys. } \textbf{\bibinfo{volume}{B275}},
\bibinfo{pages}{734} (\bibinfo{year}{1986}). 

\bibitem[{\citenamefont{Bilgici et al.}(2008)}]{Bilgici}
\bibinfo{author}{\bibfnamefont{E.}~\bibnamefont{Bilgici}}, 
\bibinfo{author}{\bibfnamefont{F.}~\bibnamefont{Bruckmann}},
\bibinfo{author}{\bibfnamefont{C.}~\bibnamefont{Gattringer}},
\bibnamefont{and}
  \bibinfo{author}{\bibfnamefont{C.}~\bibnamefont{Hagen}}, 
  \bibinfo{journal}{Phys. Rev. D} \textbf{\bibinfo{volume}{77}},
  \bibinfo{pages}{094007} (\bibinfo{year}{2008});
%
\bibinfo{author}{\bibfnamefont{E.}~\bibnamefont{Bilgici}}, 
\bibinfo{author}{\bibfnamefont{F.}~\bibnamefont{Bruckmann}},
\bibinfo{author}{\bibfnamefont{J.}~\bibnamefont{Danzer}},
\bibinfo{author}{\bibfnamefont{C.}~\bibnamefont{Gattringer}},
\bibinfo{author}{\bibfnamefont{C.}~\bibnamefont{Hagen}}, 
\bibinfo{author}{\bibfnamefont{E.}~\bibfnamefont{M.}~\bibnamefont{Ilgenfritz}}, 
\bibnamefont{and}
  \bibinfo{author}{\bibfnamefont{A.}~\bibnamefont{Maas}}, 
  \bibinfo{journal}{Few Body Syst.} \textbf{\bibinfo{volume}{47}},
  \bibinfo{pages}{125} (\bibinfo{year}{2010}).


\bibitem[{\citenamefont{Fischer et al.}(2009)}]{Fischer}
\bibinfo{author}{\bibfnamefont{C.}~\bibfnamefont{S.}~\bibnamefont{Fischer}},
\bibinfo{journal}{Phys. Rev. Lett.} \textbf{\bibinfo{volume}{103}},
\bibinfo{pages}{052003} (\bibinfo{year}{2009});
%
\bibinfo{author}{\bibfnamefont{C.}~\bibfnamefont{S.}~\bibnamefont{Fischer}} 
\bibnamefont{and}
\bibinfo{author}{\bibfnamefont{J.}~\bibfnamefont{A.}~\bibnamefont{Mueller}},  
\bibinfo{journal}{Phys. Rev. D} \textbf{\bibinfo{volume}{80}},
\bibinfo{pages}{074029} (\bibinfo{year}{2009}). 

\bibitem[{\citenamefont{Kashiwa et al}(2009)}]{Kashiwa4}
\bibinfo{author}{\bibfnamefont{K.}~\bibnamefont{Kashiwa}}, 
\bibinfo{author}{\bibfnamefont{H.}~\bibnamefont{Kouno}},
\bibnamefont{and}
\bibinfo{author}{\bibfnamefont{M.}~\bibnamefont{Yahiro}},
  \bibinfo{journal}{Phys. Rev. D} \textbf{\bibinfo{volume}{80}},
  \bibinfo{pages}{117901} (\bibinfo{year}{2009}). 

\bibitem[{\citenamefont{Braun et al.}(2009)}]{Braun}
\bibinfo{author}{\bibfnamefont{J.}~\bibnamefont{Braun}}, 
\bibinfo{author}{\bibfnamefont{L.}~\bibfnamefont{M.}~\bibnamefont{Haas}},
\bibinfo{author}{\bibfnamefont{F.}~\bibnamefont{Marhauser}}, 
\bibnamefont{and}
\bibinfo{author}{\bibfnamefont{J.}~\bibfnamefont{M.}~\bibnamefont{Pawlowski}},  
\bibinfo{journal}{Phys. Rev. Lett.} \textbf{\bibinfo{volume}{106}},
\bibinfo{pages}{022002} (\bibinfo{year}{2011}). 

\bibitem[{\citenamefont{Mukherjee et al.}(2010)}]{Mukherjee}
\bibinfo{author}{\bibfnamefont{T.}~\bibfnamefont{K.}~\bibnamefont{Mukherjee}},
\bibinfo{author}{\bibfnamefont{H.}~\bibnamefont{Chen}},
\bibnamefont{and}
\bibinfo{author}{\bibfnamefont{M.}~\bibnamefont{Huang}},  
\bibinfo{journal}{Phys. Rev. D} \textbf{\bibinfo{volume}{82}},
\bibinfo{pages}{034015} (\bibinfo{year}{2010}). 

\bibitem[{\citenamefont{Gatto et al.}(2010)}]{Gatto}
\bibinfo{author}{\bibfnamefont{R.}~\bibnamefont{Gatto}}, 
\bibnamefont{and}
\bibinfo{author}{\bibfnamefont{M.}~\bibnamefont{Ruggieri}},  
\bibinfo{journal}{Phys. Rev. D} \textbf{\bibinfo{volume}{82}},
\bibinfo{pages}{054027} (\bibinfo{year}{2010}). 

\bibitem[{\citenamefont{Mocsy et al.}(2004)}]{Mocsy}
\bibinfo{author}{\bibfnamefont{\'{A}.}~\bibnamefont{M\'{o}csy}}, 
\bibinfo{author}{\bibfnamefont{F.}~\bibnamefont{Sannino}},
\bibnamefont{and}
\bibinfo{author}{\bibfnamefont{K.}~\bibnamefont{Tuominen}},  
\bibinfo{journal}{Phys. Rev. Lett.} \textbf{\bibinfo{volume}{92}},
\bibinfo{pages}{182302} (\bibinfo{year}{2004}). 

\bibitem[{\citenamefont{Forcrand et al.}(2002)}]{Forcrand1}
\bibinfo{author}{\bibfnamefont{P.~}~\bibfnamefont{de}~\bibnamefont{Forcrand}},
\bibnamefont{and}
\bibinfo{author}{\bibfnamefont{O.}~\bibnamefont{Philipsen}},
  \bibinfo{journal}{Nucl. Phys.} \textbf{\bibinfo{volume}{B642}},
  \bibinfo{pages}{290} (\bibinfo{year}{2002});
  \bibinfo{journal}{Nucl. Phys.} \textbf{\bibinfo{volume}{B673}},
  \bibinfo{pages}{170} (\bibinfo{year}{2003}).

\bibitem[{\citenamefont{D'Elia et al.}(2003)}]{D'Elia}
\bibinfo{author}{\bibfnamefont{M.}~\bibnamefont{D'Elia}},
\bibnamefont{and}
\bibinfo{author}{\bibfnamefont{M.-P.}~\bibnamefont{Lombardo}},
  \bibinfo{journal}{Phys. Rev. D} \textbf{\bibinfo{volume}{67}},
  \bibinfo{pages}{014505} (\bibinfo{year}{2003});
  \bibinfo{journal}{Phys. Rev. D} \textbf{\bibinfo{volume}{70}},
  \bibinfo{pages}{074509} (\bibinfo{year}{2004}).

\bibitem[{\citenamefont{Chen et al.}(2005)}]{Chen}
\bibinfo{author}{\bibfnamefont{H.-S.}~\bibnamefont{Chen}},
\bibnamefont{and}
\bibinfo{author}{\bibfnamefont{X.-Q.}~\bibnamefont{Luo}},
  \bibinfo{journal}{Phys. Rev. D} \textbf{\bibinfo{volume}{72}},
  \bibinfo{pages}{034504} (\bibinfo{year}{2005}).

\bibitem[{\citenamefont{Wu et al.}(2007)}]{Wu}
\bibinfo{author}{\bibfnamefont{L.}~\bibfnamefont{K.}~\bibnamefont{Wu}},
\bibinfo{author}{\bibfnamefont{X.}~\bibfnamefont{Q.}~\bibnamefont{Luo}},
\bibnamefont{and}
\bibinfo{author}{\bibfnamefont{H.}~\bibfnamefont{S.}~\bibnamefont{Chen}},
  \bibinfo{journal}{Phys.\ Rev.\  D} \textbf{\bibinfo{volume}{76}},
  \bibinfo{pages}{034505} (\bibinfo{year}{2007}).

\bibitem[{\citenamefont{Frcrand et al.}(2010)}]{Forcrand2}
\bibinfo{author}{\bibfnamefont{P.}~\bibfnamefont{de}~\bibnamefont{Forcrand}},
\bibnamefont{and}
\bibinfo{author}{\bibfnamefont{O.}~\bibnamefont{Philipsen}},
\bibinfo{journal}{Phys. Rev. Lett.} \textbf{\bibinfo{volume}{105}},
\bibinfo{pages}{152001} (\bibinfo{year}{2010}). 

\bibitem[{\citenamefont{Frcrand et al.}(2006)}]{Forcrand3}
\bibinfo{author}{\bibfnamefont{P.}~\bibfnamefont{de}~\bibnamefont{Forcrand}},
\bibnamefont{and}
\bibinfo{author}{\bibfnamefont{S.}~\bibnamefont{Kratochvila}},
  \bibinfo{journal}{Nucl. Phys. B (Proc. Suppl.)} 
  \textbf{\bibinfo{volume}{153}},
  \bibinfo{pages}{62} (\bibinfo{year}{2006}). 

\bibitem[{\citenamefont{Li et al.}(2009)}]{Li}
\bibinfo{author}{\bibfnamefont{A.}~\bibnamefont{Li}},
\bibinfo{author}{\bibfnamefont{A.}~\bibnamefont{Alexandru}},
\bibinfo{author}{\bibfnamefont{K.-F.}~\bibnamefont{Liu}},
\bibnamefont{and}
\bibinfo{author}{\bibfnamefont{X.}~\bibfnamefont{M.}~\bibnamefont{Meng}},
  \bibinfo{journal}{Phys. Rev. D} 
  \textbf{\bibinfo{volume}{82}},
  \bibinfo{pages}{054502} (\bibinfo{year}{2010}). 

\bibitem[{\citenamefont{Cea et al.}(2010)}]{Cea1}
\bibinfo{author}{\bibfnamefont{P.}~\bibnamefont{Cea}},
\bibinfo{author}{\bibfnamefont{L.}~\bibnamefont{Cosmai}},
\bibinfo{author}{\bibfnamefont{M}~\bibnamefont{D'Elia}},
\bibnamefont{and}
\bibinfo{author}{\bibfnamefont{A.}~\bibnamefont{Papa}},
  \bibinfo{journal}{LATTICE} 
  \textbf{\bibinfo{volume}{2010}},
  \bibinfo{pages}{173} (\bibinfo{year}{2010}). 

\bibitem[{\citenamefont{Cea et al.}(2010)}]{Cea2}
\bibinfo{author}{\bibfnamefont{P.}~\bibnamefont{Cea}},
\bibinfo{author}{\bibfnamefont{L.}~\bibnamefont{Cosmai}},
\bibinfo{author}{\bibfnamefont{M}~\bibnamefont{D'Elia}},
\bibnamefont{and}
\bibinfo{author}{\bibfnamefont{A.}~\bibnamefont{Papa}},
  \bibinfo{journal}{Phys. Rev. D} 
  \textbf{\bibinfo{volume}{81}},
  \bibinfo{pages}{094502} (\bibinfo{year}{2010}). 

\bibitem[{\citenamefont{Sakai et al.}(2010)}]{Sakai3}
\bibinfo{author}{\bibfnamefont{Y.}~\bibnamefont{Sakai}},
\bibinfo{author}{\bibfnamefont{T.}~\bibnamefont{Sasaki}},
\bibinfo{author}{\bibfnamefont{H.}~\bibnamefont{Kouno}},
\bibnamefont{and}
\bibinfo{author}{\bibfnamefont{M.}~\bibnamefont{Yahiro}},
\bibinfo{journal}{Phys. Rev. D} \textbf{\bibinfo{volume}{82}},
\bibinfo{pages}{076003} (\bibinfo{year}{2010}). 

\bibitem[{\citenamefont{Contrera et al.}(2010)}]{Contrera}
\bibinfo{author}{\bibfnamefont{G.A.}~\bibnamefont{Contrera}}, 
\bibinfo{author}{\bibfnamefont{M.}~\bibnamefont{Orsaria}}, 
\bibnamefont{and}
\bibinfo{author}{\bibfnamefont{N.N.}~\bibnamefont{Scoccola}},  
\bibinfo{journal}{Phys. Rev. D} \textbf{\bibinfo{volume}{82}},
\bibinfo{pages}{054026} (\bibinfo{year}{2010}). 

\bibitem[{\citenamefont{Hell et al.}(2011)}]{Hell3}
\bibinfo{author}{\bibfnamefont{T.}~\bibnamefont{Hell}},
\bibinfo{author}{\bibfnamefont{K.}~\bibnamefont{Kashiwa}},
\bibnamefont{and}
\bibinfo{author}{\bibfnamefont{W.}~\bibnamefont{Weise}},
 \bibinfo{journal}{Phys. Rev. D} \textbf{\bibinfo{volume}{83}},
\bibinfo{pages}{114008} (\bibinfo{year}{2011}). 

\bibitem[{\citenamefont{Kondo}(2010)}]{Kondo}
\bibinfo{author}{\bibfnamefont{K.-I.}~\bibnamefont{Kondo}},
  \bibinfo{journal}{Phys. Rev. D} 
  \textbf{\bibinfo{volume}{82}},
  \bibinfo{pages}{065024} (\bibinfo{year}{2010}). 

\bibitem[{\citenamefont{Langfeld et al.}(1996)}]{Langfeld}
\bibinfo{author}{\bibfnamefont{K.}~\bibnamefont{Langfeld}}, 
\bibinfo{author}{\bibfnamefont{K.}~\bibnamefont{Kettner}}, 
\bibnamefont{and}
\bibinfo{author}{\bibfnamefont{H.}~\bibnamefont{Reinhardt}},  
  \bibinfo{journal}{Nucl. Phys.} 
  \textbf{\bibinfo{volume}{A608}},
  \bibinfo{pages}{331} (\bibinfo{year}{1996}). 

\bibitem[{\citenamefont{Nam et al.}(2010)}]{Nam}
\bibinfo{author}{\bibfnamefont{S.}~\bibfnamefont{i.}~\bibnamefont{Nam}}, 
  \bibinfo{journal}{J. Phys. \  G: Nucl. Part. Phys.} 
  \textbf{\bibinfo{volume}{37}},
  \bibinfo{pages}{075002} (\bibinfo{year}{2010}). 

\bibitem[{\citenamefont{Pagura et al.}(2011)}]{Pagura}
\bibinfo{author}{\bibfnamefont{V.}~\bibnamefont{Pagura}}, 
\bibinfo{author}{\bibfnamefont{D.}~\bibnamefont{Gomez Dumm}}, 
\bibnamefont{and}
\bibinfo{author}{\bibfnamefont{N.N.}~\bibnamefont{Scoccola}},  
\bibinfo{journal}{arXiv:hep-ph/1105.1739} (\bibinfo{year}{2011}). 



\bibitem[{\citenamefont{Roessner et al.}(2008)}]{Roessner}
\bibinfo{author}{\bibfnamefont{S.}~\bibnamefont{R\"{o}{\ss}ner}},
\bibinfo{author}{\bibfnamefont{T.}~\bibnamefont{Hell}},
\bibinfo{author}{\bibfnamefont{C.}~\bibnamefont{Ratti}},
\bibnamefont{and}
\bibinfo{author}{\bibfnamefont{W.}~\bibnamefont{Weise}},  
  \bibinfo{journal}{Nucl. Phys.} \textbf{\bibinfo{volume}{A814}},
  \bibinfo{pages}{118} (\bibinfo{year}{2008}). 

\bibitem[{\citenamefont{Cristoforetti et al.}(2009)}]{Cristoforetti}
\bibinfo{author}{\bibfnamefont{M.}~\bibnamefont{Cristoforetti}},
\bibinfo{author}{\bibfnamefont{T.}~\bibnamefont{Hell}},
\bibinfo{author}{\bibfnamefont{B.}~\bibnamefont{Klein}},
\bibnamefont{and}
\bibinfo{author}{\bibfnamefont{W.}~\bibnamefont{Weise}},  
 \bibinfo{journal}{Phys. Rev. D} \textbf{\bibinfo{volume}{81}},
  \bibinfo{pages}{114017} (\bibinfo{year}{2010}). 

\bibitem[{\citenamefont{Noguera et al.}(2008)}]{Noguera}
\bibinfo{author}{\bibfnamefont{S.}~\bibnamefont{Noguera}},
\bibnamefont{and}
\bibinfo{author}{\bibfnamefont{N.}~\bibfnamefont{N.}~\bibnamefont{Scoccola}},
  \bibinfo{journal}{Phys.\ Rev.\  D} \textbf{\bibinfo{volume}{78}},
  \bibinfo{pages}{114002} (\bibinfo{year}{2008}).

\bibitem[{\citenamefont{Bowman et al.}(2002)}]{Bowman}
\bibinfo{author}{\bibfnamefont{P.}~\bibfnamefont{O.}~\bibnamefont{Bowman}},
\bibinfo{author}{\bibfnamefont{U.}~\bibfnamefont{M.}~\bibnamefont{Heller}},
\bibnamefont{and}
\bibinfo{author}{\bibfnamefont{A.}~\bibfnamefont{G.}~\bibnamefont{Williams}},
  \bibinfo{journal}{Phys.\ Rev.\  D} \textbf{\bibinfo{volume}{66}},
  \bibinfo{pages}{014505} (\bibinfo{year}{2002}).

\bibitem[{\citenamefont{Parappilly et al.}(2006)}]{Parappilly}
\bibinfo{author}{\bibfnamefont{M.}~\bibfnamefont{B.}~\bibnamefont{Parappilly}},
\bibinfo{author}{\bibfnamefont{P.}~\bibfnamefont{O.}~\bibnamefont{Bowman}},
\bibinfo{author}{\bibfnamefont{U.}~\bibfnamefont{M.}~\bibnamefont{Heller}},
\bibinfo{author}{\bibfnamefont{D.}~\bibfnamefont{B.}~\bibnamefont{Leinweber}},
\bibinfo{author}{\bibfnamefont{A.}~\bibfnamefont{G.}~\bibnamefont{Williams}},
\bibnamefont{and}
\bibinfo{author}{\bibfnamefont{J.}~\bibfnamefont{B.}~\bibnamefont{Zhang}},
  \bibinfo{journal}{Phys.\ Rev.\  D} \textbf{\bibinfo{volume}{73}},
  \bibinfo{pages}{054504} (\bibinfo{year}{2006}).


\bibitem[{\citenamefont{Schaefer et al.}(2010)}]{Schaefer}
\bibinfo{author}{\bibfnamefont{B.-J.}~\bibnamefont{Schaefer}},
\bibinfo{author}{\bibfnamefont{J.}~\bibfnamefont{M.}~\bibnamefont{Pawlowski}},
\bibnamefont{and}
\bibinfo{author}{\bibfnamefont{J.}~\bibnamefont{Wambach}},
  \bibinfo{journal}{Phys.\ Rev.\  D} \textbf{\bibinfo{volume}{76}},
  \bibinfo{pages}{074023} (\bibinfo{year}{2007}).

\bibitem[{\citenamefont{Meif}(2005)}]{Meif}
\bibinfo{author}{\bibfnamefont{M.}~\bibnamefont{Huang}},
\bibinfo{journal}{Int. J. Mod. Phys. E} \textbf{\bibinfo{volume}{14}},
\bibinfo{pages}{675} (\bibinfo{year}{2005}).

\bibitem[{\citenamefont{Herbst et al.}(2010)}]{Herbst}
\bibinfo{author}{\bibfnamefont{T.}~\bibfnamefont{K.}~\bibnamefont{Herbst}},
\bibinfo{author}{\bibfnamefont{J.}~\bibfnamefont{M.}~\bibnamefont{Pawlowski}},
\bibnamefont{and}
\bibinfo{author}{\bibfnamefont{B.-J.}~\bibnamefont{Schaefer}},
  \bibinfo{journal}{Phys. Lett. B} \textbf{\bibinfo{volume}{696}},
  \bibinfo{pages}{58} (\bibinfo{year}{2011}). 

\bibitem[{\citenamefont{Skokov et al.}(2010)}]{Skokov}
\bibinfo{author}{\bibfnamefont{V.}~\bibnamefont{Skokov}},
\bibinfo{author}{\bibfnamefont{B.}~\bibnamefont{Friman}},
\bibnamefont{and}
\bibinfo{author}{\bibfnamefont{K}~\bibnamefont{Redlich}},
  \bibinfo{journal}{Phys. Rev. C} \textbf{\bibinfo{volume}{83}},
  \bibinfo{pages}{054904} (\bibinfo{year}{2011}). 

\bibitem[{\citenamefont{Morita et al.}(2011)}]{Morita}
\bibinfo{author}{\bibfnamefont{K.}~\bibnamefont{Morita}},
\bibinfo{author}{\bibfnamefont{V.}~\bibnamefont{Skokov}},
\bibinfo{author}{\bibfnamefont{B.}~\bibnamefont{Friman}},
\bibnamefont{and}
\bibinfo{author}{\bibfnamefont{K}~\bibnamefont{Redlich}},
\bibinfo{howpublished}{arXiv:hep-ph/1108.0735} (\bibinfo{year}{2011}).

\bibitem[{\citenamefont{Kratochvila et al.}(2006)}]{Kratochvila}
\bibinfo{author}{\bibfnamefont{S.}~\bibnamefont{Kratochvila}},
\bibnamefont{and}
\bibinfo{author}{\bibfnamefont{P.}~\bibfnamefont{de}~\bibnamefont{Forcrand}},  
  \bibinfo{journal}{Phys. Rev. D} \textbf{\bibinfo{volume}{73}},
  \bibinfo{pages}{114512} (\bibinfo{year}{2006}). 

\bibitem[{\citenamefont{D'Elia et al.}(2009)}]{D'Elia2}
\bibinfo{author}{\bibfnamefont{M.}~\bibnamefont{D'Elia}},
\bibnamefont{and}
\bibinfo{author}{\bibfnamefont{F.}~\bibnamefont{Sanfilippo}},
\bibinfo{journal}{Phys. Rev. D} \textbf{\bibinfo{volume}{80}},
\bibinfo{pages}{111501(R)} (\bibinfo{year}{2009}). 

\bibitem[{\citenamefont{Bonati et al.}(2011)}]{Bonati}
\bibinfo{author}{\bibfnamefont{C.}~\bibnamefont{Bonati}},
\bibinfo{author}{\bibfnamefont{G.}~\bibnamefont{Cossu}},
\bibinfo{author}{\bibfnamefont{M.}~\bibnamefont{D'Elia}},
\bibnamefont{and}
\bibinfo{author}{\bibfnamefont{F.}~\bibnamefont{Sanfilippo}},
\bibinfo{journal}{Phys. Rev. D} \textbf{\bibinfo{volume}{83}},
\bibinfo{pages}{054505} (\bibinfo{year}{2011}). 

\bibitem[{\citenamefont{Aarts et al.}(2010)}]{Aarts}
\bibinfo{author}{\bibfnamefont{G.}~\bibnamefont{Aarts}},
\bibinfo{author}{\bibfnamefont{S.}~\bibfnamefont{P.}~\bibnamefont{Kumar}},
\bibnamefont{and}
\bibinfo{author}{\bibfnamefont{J.}~\bibnamefont{Rafferty}},
\bibinfo{journal}{JHEP} \textbf{\bibinfo{volume}{1007}},
\bibinfo{pages}{056} (\bibinfo{year}{2010}). 


\bibitem[{\citenamefont{Barducci et al.}(1993)}]{BCPG}
\bibinfo{author}{\bibfnamefont{A.}~\bibnamefont{Barducci}}, 
\bibinfo{author}{\bibfnamefont{R.}~\bibnamefont{Casalbuoni}},
\bibinfo{author}{\bibfnamefont{G.}~\bibnamefont{Pettini}},
\bibnamefont{and}
  \bibinfo{author}{\bibfnamefont{R.}~\bibnamefont{Gatto}}, 
  \bibinfo{journal}{Phys. Lett. B} \textbf{\bibinfo{volume}{301}},
  \bibinfo{pages}{95} (\bibinfo{year}{1993}). 


\bibitem[{\citenamefont{Vogl et al.}(1991)}]{Vogl}
\bibinfo{author}{\bibfnamefont{U.}~\bibnamefont{Vogl}},
\bibnamefont{and}
\bibinfo{author}{\bibfnamefont{W.}~\bibnamefont{Weise}},
  \bibinfo{journal}{Prog. Part. Nucl. Phys.} 
  \textbf{\bibinfo{volume}{27}},
  \bibinfo{pages}{195} (\bibinfo{year}{1991}).

\bibitem[{\citenamefont{Hatsuda et al.}(1994)}]{Hatsuda}
\bibinfo{author}{\bibfnamefont{T.}~\bibnamefont{Hatsuda}},
\bibnamefont{and}
\bibinfo{author}{\bibfnamefont{T.}~\bibnamefont{Kunihiro}},
  \bibinfo{journal}{Phys.\ Rep.} \textbf{\bibinfo{volume}{247}},
  \bibinfo{pages}{221} (\bibinfo{year}{1994}).

\bibitem[{\citenamefont{Klevansky}(1992)}]{Klevansky}
\bibinfo{author}{\bibfnamefont{S.}~\bibfnamefont{P.}~\bibnamefont{Klevansky}},
  \bibinfo{journal}{Rev.\ Mod.\  Phys.} \textbf{\bibinfo{volume}{64}},
  \bibinfo{pages}{649} (\bibinfo{year}{1992}).


\bibitem[{\citenamefont{Buballa}(2005)}]{Buballa}
\bibinfo{author}{\bibfnamefont{M.}~\bibnamefont{Buballa}},
  \bibinfo{journal}{Phys.\ Rep.} \textbf{\bibinfo{volume}{407}},
  \bibinfo{pages}{205} (\bibinfo{year}{2005}).

\bibitem[{\citenamefont{Schaufer et al.}(1998)}]{Schaufer}
\bibinfo{author}{\bibfnamefont{T.}~\bibnamefont{Sch{\"a}fer}},
\bibnamefont{and}
\bibinfo{author}{\bibfnamefont{E.}~\bibnamefont{Shuryak}},
  \bibinfo{journal}{Rev.\ Mod.\ Phys.} \textbf{\bibinfo{volume}{70}},
  \bibinfo{pages}{323} (\bibinfo{year}{1998}).

\bibitem[{\citenamefont{Rapp et al.}(2000)}]{Rapp}
\bibinfo{author}{\bibfnamefont{R.}~\bibnamefont{Rapp}},
\bibinfo{author}{\bibfnamefont{T.}~\bibnamefont{Sch{\"a}fer}},
\bibinfo{author}{\bibfnamefont{E.}~\bibnamefont{Shuryak}},
\bibnamefont{and}
\bibinfo{author}{\bibfnamefont{M.}~\bibnamefont{Velkovsky}},
  \bibinfo{journal}{Ann.\ Phys.} \textbf{\bibinfo{volume}{280}},
  \bibinfo{pages}{35} (\bibinfo{year}{2000}).

\bibitem[{\citenamefont{Klimt et al.}(1991)}]{Klimt}
\bibinfo{author}{\bibfnamefont{S.}~\bibnamefont{Klimt}},
\bibinfo{author}{\bibfnamefont{M.}~\bibnamefont{Lutz}},
\bibinfo{author}{\bibfnamefont{U.}~\bibnamefont{Vogl}},
\bibnamefont{and}
\bibinfo{author}{\bibfnamefont{W.}~\bibnamefont{Weise}},
\bibinfo{journal}{Nucl.\ Phys.} \textbf{\bibinfo{volume}{A516}},
  \bibinfo{pages}{429} (\bibinfo{year}{1990}). 


\bibitem[{\citenamefont{Frank et al.}(2003)}]{Frank}
\bibinfo{author}{\bibfnamefont{M.}~\bibnamefont{Frank}},
\bibinfo{author}{\bibfnamefont{M.}~\bibnamefont{Buballa}},
\bibnamefont{and}
\bibinfo{author}{\bibfnamefont{M.}~\bibnamefont{Oertel}},
  \bibinfo{journal}{Phys.\ Lett.\  B} \textbf{\bibinfo{volume}{562}},
  \bibinfo{pages}{221} (\bibinfo{year}{2003}).

\bibitem[{\citenamefont{Kunihiro}(1989)}]{Kunihiro}
\bibinfo{author}{\bibfnamefont{T.}~\bibnamefont{Kunihiro}},
  \bibinfo{journal}{Phys.\ Lett.\  B} \textbf{\bibinfo{volume}{219}},
  \bibinfo{pages}{363} (\bibinfo{year}{1989}).


\bibitem[{\citenamefont{Rehberg et al.}(1996)}]{Rehberg}
\bibinfo{author}{\bibfnamefont{P.}~\bibnamefont{Rehberg}},
\bibinfo{author}{\bibfnamefont{S.}~\bibfnamefont{P.}~\bibnamefont{Klevansky}},
\bibnamefont{and}
\bibinfo{author}{\bibfnamefont{J.}~\bibnamefont{H{\"u}fner}},
  \bibinfo{journal}{Phys.\ Rev.\  C} \textbf{\bibinfo{volume}{53}},
  \bibinfo{pages}{410} (\bibinfo{year}{1996}).


\end{thebibliography}
\end{document}